\documentclass[12pt]{article}
\usepackage[a4paper, top=16mm, text={170mm, 248mm}, includehead, includefoot, hmarginratio=1:1, heightrounded]{geometry}
\usepackage{amsmath,amssymb,mathrsfs,amsthm,tikz,shuffle,paralist}
\usepackage{dsfont}
\usepackage{color}
\definecolor{darkred}{RGB}{173,34,48}

\usetikzlibrary{snakes}
\usetikzlibrary{calc}
\usetikzlibrary{decorations}
\usepackage[all]{xypic}
\usepackage{jheppub}
\usepackage{hyperref}
\usepackage{subfigure}
\usepackage{caption}
\usepackage{tikz}
\usepackage{subfigure,tikz}
\usetikzlibrary{backgrounds,automata}
\newcommand{\bea}{\begin{eqnarray}}
\newcommand{\eea}{\end{eqnarray}}
\newcommand{\bean}{\begin{eqnarray*}}
\newcommand{\eean}{\end{eqnarray*}}
\newcommand{\nn}{\nonumber \\}

\def\no{\nonumber}

\def\W #1{\widetilde{#1}}

\def\eref#1{(\ref{#1})}

\def\d{{\rm d}}

\def\a{{\alpha}}

\def\b{{\beta}}

\def\d{\partial}

\def\eps{\epsilon}

\def\Label#1{\label{#1}%
  \smash{\hbox to0pt{\raise1ex\hbox{\tiny[#1]}\hss}}}

\preprint{USTC-ICTS/PCFT-21-26 }

\title{\boldmath Analytic Tadpole Coefficients of One-loop Integrals}

\author[a,b,c,e]{Bo Feng,}
\author[a,1]{Tingfei Li,\note{Corresponding author.}}
\author[d,e]{Xiaodi Li}

\affiliation[a]{Zhejiang Institute of Modern Physics, Zhejiang University, Hangzhou, 310027, P. R. China }
\affiliation[b]{Beijing Computational Science Research Center, Beijing 100084, China}
\affiliation[c]{Center of Mathematical Science, Zhejiang University, Hangzhou, 310027, P. R. China}
\affiliation[d]{Interdisciplinary Center for Theoretical Study, University of Science and Technology of China, Hefei, Anhui 230026, China}
\affiliation[e]{Peng Huanwu Center for Fundamental Theory, Hefei, Anhui 230026, China}

\emailAdd{fengbo@zju.edu.cn}
\emailAdd{tfli@zju.edu.cn}
\emailAdd{lixiaodi@zju.edu.cn}

\abstract{One remaining problem of unitarity cut method for one-loop integral reduction is that
 tadpole coefficients can not be straightforward obtained through this way.  In this paper, we reconsider the problem by applying differential operators over an auxiliary vector $R$.
Using differential operators, we establish the corresponding differential equations for tadpole
coefficients at the first step. Then using the tensor structure of tadpole coefficients, we transform the differential
equations to the recurrence relations for undetermined tensor coefficients. These recurrence
relations can be solved easily by iteration and we can obtain analytic expressions of tadpole coefficients for arbitrary one-loop integrals.

}


\begin{document}
\maketitle
\flushbottom

\section{Introduction}
% % % % % % % % %
Recent decades have witnessed much progress in the calculation of higher loop scattering amplitudes. Among them, one of the most remarkable achievements is the almost completely solution of  the calculation of the one loop amplitude by many tools, such as  PV reduction\cite{Passarino:1978jh}, OPP reduction\cite{Ossola_2007}, Unitarity cut \cite{Bern:1994zx,Bern:1994cg,Britto:2004nc} etc..

It's well known that \cite{Passarino:1978jh, Ossola_2007, Bern:1994cg} under dimensional regularization, an one-loop scattering amplitude or a general one-loop Feynman integral in $D=d_0-2\epsilon$-dimensional spacetime can be rewritten as a linear combination of some scalar master integrals (MIs) as
\begin{align}
 I=\sum_{i_{d_0+1}} C_{d_0+1}^{i_{d_0+1}} I_{d_0+1}^{i_{d_0+1}}+\sum_{i_{d_0}} C_{d_0}^{i_{d_0}} I_{d_0}^{i_{d_0}}+\cdots + \sum_{i_1} C_1^{i_1} I_1^{i_1}, \label{reduction_d0}
\end{align}
where the coefficient $C_s^{i_s}$ ($s=1,\cdots,d_0+1$) is a rational function of external momenta, and $I_s^{i_s}$ is the $s$-gon scalar integral. If $d_0=4$, then the corresponding master integrals are traditionally referred as tadpole, bubble, triangle, box and pentagon integrals (the tadpole integrals vanish if the corresponding propagators are massless). 
So the task of computing an one-loop amplitude is reduced to determining the coefficients of these master integrals.
%So the task of computation of one-loop amplitude is reduced to the determination of the coefficients of these master integrals.
These coefficients can be derived by either integrand reduction \cite{Ossola_2007} algebraically, or the unitarity method \cite{Bern:1994zx,Bern:1994cg,Britto:2004nc}.
The idea underlying the unitarity method is to constrain amplitudes by their branch cuts, more explicitly, comparing two sides of \eref{reduction_d0} after cutting several propagators.
By using the usual unitarity cuts (cut two propagators corresponding to a physical channel), the explicitly analytical expressions of these coefficients have been given in  \cite{Britto:2006fc,Britto:2007tt,Britto:2008vq}.
However, the analytic results of the tadpole coefficients are missing.

To obtain the coefficients of tadpole integrals by the usual unitary cut method, an idea was proposed to add an auxiliary, unphysical propagator in the integrand \cite{Britto:2009wz}. Since we need to compare the reduction coefficients of the physical integrand and auxiliary integrand 
in this new frame, the method isn't very efficient to calculate tadpole coefficients.
Another idea is that although the tadpole integrals vanish under the unitarity cuts in physical channels, it survives under the single cut. Then naturally the single cut method was used to calculate the tadpole coefficients \cite{Britto:2010um}.
Because of the divergence of integrals after single cut and the dependence on the tensor reduction of the integrand, it's not easy to compute the tadpole coefficients for a general case.
%Because of the divergence of the single cut integrals and the dependence on the tensor reduction of the integrand, it's not easy to compute the tadpole coefficients for a general case.

In this paper, we will reconsider the  computation of tadpole coefficients by using   differential operators.  Differential operators have played an important role in the area of scattering amplitude, for example, deriving the IBP relations and differential equations of Feynman integrals \cite{Smirnov:2012gma}, relating tree-level amplitudes of different theories \cite{Cheung:2017ems} and the expansion of Einstein-Yang-Mills amplitude \cite{Feng:2019tvb, Feng:2020jck}.

Roughly speaking, for a general tensor one-loop integral, we will first introduce an auxiliary vector $R^{\mu}$ and assume its reduction to scalar master integrals\footnote{The auxiliary vector $R^{\mu}$ closely resembles the polarization vectors in the expansion of Einstein-Yang-Mills amplitude \cite{Feng:2019tvb, Feng:2020jck}.}, then consider applying some differential operators with respect to $R$ to the integral, so we obtain the differential equations of tadpole coefficients after comparing two sides of  the equation. Actually, the action of these differential operators of $R$ is a little similar to the traditional PV reduction \cite{Passarino:1978jh}. Instead of trying to solve these differential equations directly, we transform the differential equations into recurrence relations with the help of the general tensor form of tadpole coefficients. With some known initial conditions, we can solve these expansion coefficients iteratively. So the problem of calculating  tadpole coefficients is reduced into solving these recurrence relations, we will provide a general algorithm.

Our plan of this paper is following.
In section \ref{sec:recurrence relation}, we discuss  the integral reduction of a general tensor $1$-loop Feynman integral.  First, we consider the action of differential operators and obtain the differential equations of reduction coefficients. Second, we transform these differential equations into recurrence relations.
In section \ref{sec:tadpole}, we derive the recurrence relations of tadpole coefficients for four tensor integrals, namely bubbles, triangles, boxes, pentagons, and provide a general algorithm for calculating tadpole coefficients with some examples.
Appendix \ref{sec:PVreduction} briefly reviews the traditional PV-reduction method to deal with the tadpole coefficient of a tensor tadpole, which is consistent with our results.

%%%%%%%%%%%%%%%%%%
\section{Integral reduction by differential operators}
%%%%%%%%%%%%%%%%%%%%
\label{sec:recurrence relation}
In this section, we will consider the integral reduction of a general $1$-loop tensor integral using  differential operators with respect to an auxiliary vector $R$. First, in subsection \ref{sec:differential_equation} we will derive the differential equations for the coefficients of the master integrals. Secondly, using  the general tensor structure  of reduction coefficients,  we transform the differential equations of tadpole coefficients into recurrence relations in subsection \ref{sec:recurrence_relation}.

%%%%%%%%%%%%%%%%%%%%%%%%%
\subsection{Differential equations of reduction coefficients}
\label{sec:differential_equation}
%%%%%%%%%%%%%%%%%%%%%%%%%%

%In this subsection we will introduce an auxiliary vector $R^{\mu}$ and drive the differential equations under the action of corresponding differential operators.
Let us start with  the following general one-loop \textit{$m$-rank} tensor integral with $n+1$ propagators
\bea
I^{\mu_1\cdots \mu_m}_{n+1}
= \int {d^D\ell\over (2\pi)^D} \frac{\ell^{\mu_1}\ell^{\mu_2}\cdots \ell^{\mu_m}}{P_0P_1\cdots P_{n}},~\label{Idea-1-1-0}
\eea
where the $i$-th propagator is given by $P_i=(\ell-K_i)^2-M_i^2$ and $K_0=0$ (i.e., we have chosen to translate the loop momentum $\ell \to \ell+K_0$ to simplify the $0$-th propagator $P_0$).
To simplify the manipulation of the tensor structure and utilize the tool of differential operators, we introduce an auxiliary vector $R^{\mu}$ and contract $I^{\mu_1\cdots \mu_m}_{n+1}$ with $m$ $R^{\mu}$'s to arrive
\begin{align}
I^{(m)}_{n+1}[R]
\equiv& 2^m I^{\mu_1\cdots \mu_m}_{n+1} R_{\mu_1}\cdots R_{\mu_m}
=\int {d^D\ell\over (2\pi)^D} { (2\ell\cdot R)^m\over P_0P_1\cdots P_n},~\label{Idea-1-1}
\end{align}
Note that the auxiliary vector $R^{\mu}$ is in the $D=(4-2\eps)$-dimensional space as the $D$-dimensional loop momenta $\ell$. By setting $R=\sum_{i=1}^m \a_i R_i$ into \eref{Idea-1-1} and expanding the result to find the coefficients of $\a_1...\a_m$,
it is easy to see that we will get  the reduction of
\eref{Idea-1-1-0} up to a numerical factor.
%First we want to explain  that knowing  analytic expressions of the reduction of \eref{Idea-1-1}, we will know the reduction of  general one-loop integrals.
%For example, if the numerator is  $(2\ell\cdot R_1)(2\ell\cdot R_2)$, we can consider the reduction of $(2\ell\cdot R)^2$, then put $R=\a_1 R_1+\a_2 R_2$ and expand it, thus the coefficient of $2\a_1\a_2$ is the wanted reduction result for the original numerator. Similarly, if the numerator is  $4 \ell_\mu F^{\mu\nu} \ell_\nu$, we can consider the reduction of $(2\ell\cdot R_1)(2\ell\cdot R_2)$ first. Then for each pair of $R_1, R_2$, we replace $(K_1\cdot R)(K_2\cdot R)$ by $ (K_1)_\mu F^{\mu\nu} (K_2)_\nu$ and $R_1\cdot R_2$ by $\eta_{\mu\nu} F^{\mu\nu}$ (please notice that  $F^{\mu\nu}$ is symmetric tensor). By similar arguments, we see that the reduction of general tensor  \eref{Idea-1-1-0} can be replaced by the reduction of  \eref{Idea-1-1}.
The simple but useful transformation from the form \eref{Idea-1-1-0} to the form \eref{Idea-1-1} is, in fact, our first crucial step.
%To interpret the form of \eref{Idea-1-1}, we can take another point of view. For example, if the
%numerator is  $(2\ell\cdot R_1)(2\ell\cdot R_2)$, we can consider the reduction of $(2\ell\cdot R)^2$, then put $R=\a_1 R_1+\a_2 R_2$ and expand it, thus the coefficient of $2\a_1\a_2$ is the
%wanted reduction result for the original numerator. Similarly, if the
%numerator is  $4 \ell_\mu F^{\mu\nu} \ell_\nu$, we can consider the reduction of $(2\ell\cdot R_1)(2\ell\cdot R_2)$ first. Then for each pair of $R_1, R_2$, we replace $(K_1\cdot R)(K_2\cdot R)$ by
%$ (K_1)_\mu F^{\mu\nu} (K_2)_\nu$ and $R_1\cdot R_2$ by $\eta_{\mu\nu} F^{\mu\nu}$ (please notice that  $F^{\mu\nu}$ is symmetric tensor).

It's well known that in dimensional regularization scheme, the integral $I^{(m)}_{n+1}[R]$ can be reduced into the linear combination of master integrals (including pentagon, box, triangle, bubble and tadpole scalar integrals) as
\begin{align}
I^{(m)}_{n+1}[R]= \sum_{i_5} C_5^{i_5}(m) I_5^{i_5}+\sum_{i_4} C_4^{i_4}(m) I_4^{i_4}+\cdots + \sum_{i_1} C_1^{i_1}(m) I_1^{i_1},  ~~~~ \label{reduction-formula}
\end{align}
where the reduction coefficients $C_s^{i_s}(m), s=1,\cdots,5$ are rational functions of external momenta, masses and $R$. An important point of the reduction is that $R$ can only appear in the numerator of the reduction coefficients. The Lorentz invariance means that
it can only have following types of contractions:
%From the equation (\ref{Idea-1-1}), it's easy to see that the tensor structure of a reduction coefficient actually is
%
\bea
 R\cdot R;~~~~~~~~~~ R\cdot K_i, i=1,...,n;~~~\label{coefficient-tensor-structure}
\eea
%
%then we can infer that $R$ only appears in the numerator of $C_s^i(m)$. The problem of integral reduction of  $I^{(m)}_{n+1}$ is just the determination of these reduction coefficients  $C_s^i(m)$.

To find these reduction coefficients  $C_s^{i_s}(m)$, we will try to establish some differential equations by virtue of the following \textit{differential operators}:
\bea
\mathcal{D}_i\equiv K_i\cdot {\d \over \d R},~~i=1,...,n;~~~~ \quad \mathcal{T}\equiv \eta^{\mu\nu}{\d\over \d R^\mu}
{\d \over \d R^\nu}.\label{def-diffe}
 \eea
Let us start with the action of $\mathcal{D}_i$. Applying it to the equation \eref{reduction-formula}, on the left-hand side we have,
\begin{align}
\mathcal{D}_i I^{(m)}_{n+1}[R]
& =  \int {d^D\ell\over (2\pi)^D} {m (2\ell\cdot R)^{m-1} (2\ell\cdot K_i)\over (\ell^2-M_0^2) \prod_{j=1}^n
[(\ell-K_j)^2-M_j^2]} \nn
& =  \int {d^D\ell\over (2\pi)^D} {m (2\ell\cdot R)^{m-1} [ (\ell^2-M_0^2)- ((\ell-K_i)^2-M_i^2)+(M_0^2+K_i^2-M_i^2)]\over (\ell^2-M_0^2) \prod_{j=1}^n
[(\ell-K_j)^2-M_j^2]} \nn
& =  m I^{(m-1)}_{n+1;\widehat{0}}- m I^{(m-1)}_{n+1;\widehat{i}}+ m f_i I^{(m-1)}_{n+1},~~~\label{D-act-left}
\end{align}
where we  have defined $f_i\equiv M_0^2+K_i^2- M_i^2$ and the subscript $\widehat{i}$ in $I^{(m-1)}_{n+1;\widehat{i}}$ means that the propagator $P_i$ is removed from $I^{(m-1)}_{n+1}$.
On the right-hand side, since $\mathcal{D}_i$  acts only on the coefficients $C_s^{i_s}(m)$,
we have $\sum_{s=1}^5 \sum_{i_s} \left(\mathcal{D}_{i}C_s^{i_s}(m) \right) I_s^{i_s} $. Identifying both sides, 
 we get the  equation
\begin{align}
m I^{(m-1)}_{n+1;\widehat{0}}- m I^{(m-1)}_{n+1;\widehat{i}}+ m f_i I^{(m-1)}_{n+1}=\sum_{s=1}^5 \sum_{i_s} \left(\mathcal{D}_{i}C_s^{i_s}(m) \right) I_s^{i_s}. ~~~\label{Di-equation-1}
\end{align}
%
%If we put the reduction formula \eref{reduction_formula} at the left hand side and compare
%coefficients of basis, we will get the differential equations for reduction coefficients.
At this point, the differential operator $\mathcal{D}_i$ doesn't help us much for calculating the reduction coefficients. However, if we make the \textit{inductive assumption} that the reduction of tensor integral $I^{(m')}_{n'+1}$'s are already known for $m'< m, n'<n$, or $m'<m,n'=n$ and $m'=m, n'<n$, for example
\begin{align}
 I^{(m-1)}_{n+1;\widehat{0}}=\sum_{s=1}^5  \sum_{i_s} C_s^{i_s}(m-1;\widehat{0}) I_{s;\widehat{0}}^{i_s},
\end{align}
where $\widehat{0}$ of $C_s^{i_s}(m-1;\widehat{0})$ is to remind us $C_s^{i_s}(m-1;\widehat{0})$ is the reduction coefficient of an integral with propagator $P_0$ being canceled out,
the \eref{Di-equation-1} can be written as
\begin{align}
\sum_{s=1}^5 \sum_{i_s} \left(\mathcal{D}_{i}C_s^{i_s}(m) \right) I_s^{i_s}=&m \sum_{s=1}^5  \sum_{i_s} C_s^{i_s}(m-1;\widehat{0}) I_{s;\widehat{0}}^{i_s}  - m \sum_{s=1}^5  \sum_{i_s} C_s^{i_s}(m-1;\widehat{i}) I_{s;\widehat{i}}^{i_s}  \notag\\
&+ m f_i \sum_{s=1}^5  \sum_{i_s} C_s^{i_s}(m-1) I_s^{i_s}. \label{Di_equation_2}
\end{align}
By comparing the master integrals at the two sides of the above formula, we can get the differential equation for each particular reduction coefficient $C_s^{i_s}(m)$.

It is found that the differential equations given by $\mathcal{D}_i$'s are not enough to uniquely determine  the reduction coefficients, so we need to consider the action of $\mathcal{T}$ on \eref{reduction-formula}. Similar to $\mathcal{D}_i$'s, on the left-hand side, we get
\begin{align}
\mathcal{T} I^{(m)}_{n+1}
& =  4m (m-1)M_0^2 I^{(m-2)}_{n+1}+ 4m(m-1)I^{(m-2)}_{n+1;\widehat{0} },
~~~\label{Idea-2-1}
\end{align}
while on the right-hand side, we have
$\sum_{s=1}^5 \sum_{i_s} \left(\mathcal{T}C_s^{i_s}(m) \right) I_s^{i_s}$.
So after the reduction of integrals on the left-hand side, we can get another group of  differential equations for unknown reduction coefficients
\begin{align}
\sum_{s=1}^5 \sum_{i_s} \left(\mathcal{T}C_s^{i_s}(m) \right) I_s^{i_s}
=& 4m (m-1) \sum_{s=1}^5  \sum_{i_s} C_s^{i_s}(m-2;\widehat{0}) I_{s;\widehat{0}}^{i_s}   \notag\\
&+ 4m(m-1) M_0^2 \sum_{s=1}^5  \sum_{i_s} C_s^{i_s}(m-2) I_s^{i_s}.~~~ \label{T-equation}
\end{align}

\subsection{Recurrence relations for tadpole coefficients}
\label{sec:recurrence_relation}
Since our main goal is to compute the reduction coefficients of tadpole  integral,  we will concentrate on the calculation of tadpole coefficients here.
Without loss of generality, let's consider the tadpole integral with a propagator $P_0$\footnote{Knowing it, by proper loop momentum shifting, for example, $\ell\to \W\ell=\ell+K_i$, we can get the tadpole coefficients of $P_i$.}.
Comparing the two sides of equations \eref{Di_equation_2} and \eref{T-equation}, since the $0$-th propagator $P_0$ has been removed from $I_{s;\widehat{0}}^{i_s}$,  it won't contribute to the tadpole coefficient of propagator $P_0$, so we get the following differential equations
\begin{align}
\mathcal{D}_{i}C_1^{(0)}(m) = - m   C_1^{(0)}(m-1;\widehat{i}) + m f_i  C_1^{(0)}(m-1),   \label{tadpole_DE_1}
\end{align}
and
\bea
\mathcal{T}  C_1^{(0)}(m)= 4m (m-1)M_0^2 C_1^{(0)}(m-2),  \label{tadpole_DE_2}
\eea
where the superscript of $C_1^{(0)}(m)$  reminds us it's the tadpole coefficient with a propagator $P_0$.

Note that the above two differential equations \eref{tadpole_DE_1} and \eref{tadpole_DE_2} relate tadpole coefficient $C_1^{(0)}(m)$ with rank $m$ to tadpole coefficients with lower ranks $(m-1)$ or $(m-2)$, which are already known according to the inductive assumption.
Since directly solving the differential equations \eref{tadpole_DE_1} and \eref{tadpole_DE_2} are complicated,  we will try to transform the differential equations into much simpler recurrence relations by noticing the  tensor structure of the tadpole coefficient, i.e., it can be expanded as 
\begin{align}
C_1^{(0)}(m)= \sum^{\ \ \ \ \ \prime}_{\{i_0,\cdots,i_n\}=0} c^{(m)}_{i_0,i_1,\cdots,i_n} (M_0^2)^{i_0-n} (R\cdot R)^{i_0}\prod_{k=1}^n (R\cdot K_k)^{i_k} \label{coefficient_expansion_1}
\end{align}
with unknown coefficients $c^{(m)}_{i_0,i_1,i_2,i_3,...i_n}$ (called \textit{expansion coefficient}) being rational functions of $(K_i\cdot K_j), M_i^2$, and the  summing indices satisfying $2i_0+\sum_{k=1}^n i_k=m$ (so the summing is written as $\sum^\prime$ to emphasize the constraint and if we choose  $i_1,i_2,\cdots,i_n$ as free indices, we can just write $c^{(m)}_{i_0,i_1,\cdots, i_n}$ as $c^{(m)}_{i_1,i_2,\cdots, i_n}$ ).  The above formula \eref{coefficient_expansion_1} always exists because of the previous mentioned contractions \eref{coefficient-tensor-structure}.
For simplicity, we will also adopt the following conventions:
\begin{itemize}

\item First, we define the notations $s_{00}=R\cdot R, s_{0i}=R\cdot K_i$, for $ i=1,...,n$, and $s_{ij}=K_{i}\cdot K_j$, for $i,j=1,2,...,n$.
	
\item Second, the mass dimension of $C_1^{(0)}(m)$ is $2(m-n)$, and to make all expansion coefficients $c^{(m)}_{i_1,i_2,i_3,...i_n}$  dimensionless, we extract their mass dimension $(M_0^2)^{i_0-n}$ explicitly.
	
\item Thirdly, we choose to extend the definition domain of $i_k, k=0,1,...,n$ to $\mathbb{Z}$, but keep in mind that $c^{(m)}_{i_1,i_2,...,i_n}$ vanishes if one index $i_k, k=0,1,...,n$ is negative or $m-\sum_{k=1}^n i_k$ is odd. The convention of indices will simplify the considerations of boundary conditions of recurrence relations in the future.
\item Fourthly, due to any a permutation $\sigma : \{K_i,M_i\} \to \{K_{\sigma(i)},M_{\sigma(i)}\}$ leads both $I_{n+1}^{(m)}$ and the tadpole integral $I_{1}$ with propagator $P_0$  invariant, we have  $c^{(m)}_{i_1,\cdots,i_n}=\sigma c^{(m)}_{i_{\sigma^{-1}(1)},\cdots,i_{\sigma^{-1}(n)}} $. The permutation symmetry of the expansion coefficients can be used to check our results.
\end{itemize}
After taking  above notations, the tadpole coefficient in \eref{coefficient_expansion_1} becomes
\begin{align}
C_1^{(0)}(m)
= & \sum^{\ \ \ \ \ \prime}_{\{i_0,\cdots,i_n\} } c^{(m)}_{i_0,i_1,\cdots,i_n}  (M_0^2)^{i_0-n} s_{00}^{i_0}\prod_{k=1}^n s_{0k}^{i_k},~~~~~\label{C0-exp-1}
\end{align}
and our goal is to determine the unknown expansion coefficients $c^{(m)}_{i_0,i_1,\cdots,i_n}$.
When using above notation to express $C_1^{(0)}(m-1;\widehat{i})$,
one need to notice that it doesn't contain the external momenta $K_i$  for the propagator $P_i$ has been canceled in integral $I_{s;\widehat{i}}^{i_s}$, so its expansion is
\begin{align}
C_1^{(0)}(m-1;\widehat{i})
=& \sum^{\ \ \ \ \ \prime}_{\{j_0,\cdots,j_n\} }   c^{(m-1)}_{j_0,\cdots,\widehat{j}_i,\cdots,j_n}[\widehat{i}] (M_0^2)^{j_0-(n-1)} s_{00}^{j_0}\prod_{k=1, k\ne i}^n s_{0k}^{j_k} \notag\\
=&  \sum^{\ \ \ \ \ \prime}_{\{j_0,\cdots,j_n\}}   \delta_{0j_i}~   c^{(m-1)}_{j_0,\cdots,\widehat{j}_i,\cdots,j_n}[\widehat{i}] (M_0^2)^{j_0-(n-1)}s_{00}^{j_0}\prod_{k=1}^n s_{0k}^{j_k} ,  \label{coefficient_expansion_2}
\end{align}
where in the second equation, we have added $\delta_{0j_i}$ and the summation over $j_i$ for later convenience.

To get the recurrence relations, we need to consider the action of $\mathcal{D}_i, \mathcal{T}$ on the \eref{C0-exp-1}. Using the chain rule, it's easy to get
 \begin{align}
 \mathcal{D}_i
 =K_i^{\mu}{\d \over \d R^{\mu}}=2s_{0i}{\d \over \d s_{00}}+\sum_{j=1}^ns_{ij}{\d \over \d s_{0j}},   ~~~~\label{Di-Exp}
 \end{align}
and
 \bea
 \mathcal{T}
 =2D{\d \over \d s_{00}}+4s_{00}{\d^2\over \d s_{00}^2}+4\sum_{i=1}^ns_{0i}{\d \over \d s_{0i}}{\d \over \d s_{00}}+\sum_{i=1}^n\sum_{j=1}^ns_{ij}{\d \over \d s_{0i}}{\d \over \d s_{0j}},  ~~~~\label{T-Exp}
 \eea
 where $D$ is the space-time dimension.
 First, we consider the action of differential operator $\mathcal{D}_j$ in \eref{tadpole_DE_1}
 \begin{align}
 \mathcal{D}_jC_1^{(0)}(m)
 =&\sum^{\ \ \ \ \  \prime}_{i_0,i_1,...,i_n}c^{(m)}_{i_0,i_1,...,i_n}(M_0^2)^{i_0-n} \prod_{k=0}^n s_{0k}^{i_k}\left[ 2i_0 s_{0j} s_{00}^{-1}+\sum_{l=1}^n i_ls_{jl}s_{0l}^{-1}\right]\nn
 =&\sum^{\ \ \ \ \ \prime}_{i_1,i_2,...,i_n}(M_0^2)^{(m-1-\tilde{i})/ 2-n}s_{00}^{(m-1-\tilde{i})/2}\prod_{k=1}^n s_{0k}^{i_k}\nn
 &\left[  (m+1-\sum_{l=1}^ni_l)M_0^2 c^{(m)}_{i_1,\cdots,i_j-1,\cdots,i_n}
 +\sum_{l=1}^n(i_l+1)s_{jl}c^{(m)}_{i_1,\cdots,i_l+1,\cdots,i_n}\right],
 \end{align}
 where in the second equation we have used the fact $2i_0+\sum_{k=1}^n i_k=m$ and redefined the indices $i_j$ and $i_l$, and $\tilde{i}\equiv \sum_{k=1}^n i_k$.
Using the expansion \eref{C0-exp-1},  the right-hand side is
 \begin{align}
 &mf_jC_1^{(0)}(m-1)-mC_1^{(0)}(m-1;\widehat{j}) \notag\\
 =&\sum_{i_1,\cdots,i_n}c^{(m-1)}_{i_1,\cdots,i_n}  mf_j   (M_0^2)^{(m-1-\widetilde{i})/2-n}   s_{00}^{(m-1-\widetilde{i})/2}\prod_{k=1}^n s_{0k}^{i_k} \notag\\
 &+ \sum_{i_1,\cdots,i_n}c^{(m-1)}_{i_1,\cdots,\widehat{i}_j,\cdots,i_n}[\widehat{j}] m M_0^2 \delta_{0 i_j}  (M_0^2)^{(m-1-\widetilde{i})/2-n}   s_{00}^{(m-1-\widetilde{i})/2}\prod_{k=1}^n s_{0k}^{i_k}.
 \end{align}
 Comparing the two sides we have
\begin{align}
&(m+1-\sum_{l=1}^ni_l) c^{(m)}_{i_1,\cdots,i_j-1,\cdots,i_n}
+\sum_{l=1}^n(i_l+1)\beta_{jl}c^{(m)}_{i_1,\cdots,i_l+1,\cdots,i_n} \notag\\
= &  m\alpha_j c^{(m-1)}_{i_1,\cdots,i_n}- m  \delta_{0 i_j}  c^{(m-1)}_{i_1,\cdots, \widehat{i}_j,\cdots, i_n}[\widehat{j}] \label{recurrence_relation_1}
\end{align}
 with $\alpha_j\equiv f_j/M_0^2, \beta_{jl}\equiv s_{jl}/M_0^2$.
 Secondly considering the operator $\mathcal{T}$, we have
 \begin{align}
 \mathcal{T}C_1^{(0)}(m)
 =&\sum^{\prime}_{i_0,\cdots ,i_n}c^{(m)}_{i_0,\cdots ,i_n}  (M_0^2)^{i_0-n}  \prod_{k=0}^n s_{0k}^{i_k}\nn
 &\left[2i_0(2m+D-2i_0-2)s_{00}^{-1}+\sum_{0<j<k}2i_ji_ks_{jk} s_{0j}^{-1}s_{0k}^{-1}+\sum_{j=1}^n (i_j-1)i_js_{jj} s_{0i_j}^{-2} \right]\nn
 =&\sum_{i_1,i_2,...,i_n}(M_0^2)^{(m-\tilde{i})/ 2-1-n}   s_{00}^{(m-\tilde{i})/2-1}\prod_{k=1}^n s_{0k}^{i_k}\nn
 &\Big[2i_0(2m+D-2i_0-2)M_0^2c^{(m)}_{i_1,\cdots,i_n}
 +\sum_{0<j<k}2(i_j+1)(i_k+1)s_{jk}c^{(m)}_{i_1,\cdots,i_j+1,\cdots,i_k+1,\cdots,i_n}\nn
 &+\sum_{j=1}^n(i_j+1)(i_j+2)s_{jj}c^{(m)}_{i_1,\cdots,i_j+2,\cdots,i_n} \Big],
 \end{align}
at the left hand side  and
 \bea
 4m(m-1)M_0^2\sum_{i_1,i_2,...,i_n} c^{(m-2)}_{i_1,\cdots,i_n}   (M_0^2)^{(m-\tilde{i})/ 2-1-n}   s_{00}^{(m-\tilde{i})/ 2-1}  \prod_{k=1}^n s_{0k}^{i_k}.
 \eea
at the right-hand side. Comparing two sides, we get
 \begin{align}
 4m(m-1)c^{(m-2)}_{i_1,\cdots,i_n}
 =& (m-\sum_{k=1}^ni_k)(D+m+\sum_{k=1}^ni_k-2)c^{(m)}_{i_1,\cdots,i_n}  \notag\\
 &+\sum_{0<j<k}2(i_j+1)(i_k+1)\b_{jk}c^{(m)}_{i_1,\cdots,i_j+1,\cdots,i_k+1,\cdots,i_n} \notag\\
 &+\sum_{j=1}^n(i_j+1)(i_j+2)\b_{jj}c^{(m)}_{i_1,\cdots,i_j+2,\cdots,i_n}. \label{recurrence_relation_2}
 \end{align}
Up to now, we have successfully transformed the differential equations to algebraic  recurrence relations \eref{recurrence_relation_1} and \eref{recurrence_relation_2} for tadpole coefficients, then the next step is to solve these relations  explicitly.

\section{Tadpole coefficients of tensor  integrals}
\label{sec:tadpole}
In the previous section, we have derived two types of recurrence relations (we will call them
${\cal D}$-type and ${\cal T}$-type) of unknown expansion coefficients appearing in \eref{C0-exp-1}.
In this section, we will show that by the \textit{inductive assumption}
the expansion coefficients with larger indices $i_k$ are related to those expansion coefficients with lower indices through the recurrence relations.
%by the  induction the expansion coefficients with larger indices $i_k$ are related to those expansion coefficients with lower indices by the recurrence relations.
Then with the boundary conditions\footnote{Specially, for $m=0,n\le 4$, the tensor integral $I_{n+1}^{(m)}$ is just the master integral, so it can't be reduced further, then we have $C_1^{(0)}(0)=\delta_{0,n}$}, these recurrence relations are enough to recursively determine  the tadpole coefficients for a general tensor $1$-loop integral with any rank.
To illustrate this, we will take the tensor $1$-loop integrals with up to $5$ propagators as examples in next subsections, and an explicitly recursive algorithm is provided in each subsection.

Before moving on complicated examples, let us first apply our method to the simplest example, i.e., the reduction of tensor tadpole integrals. With $n=0$ in \eref{Idea-1-1}, it is easy to see  that
the tadpole coefficients are nonzero only when $m$ is even, because the only existing Lorentz contraction is $R^2$. When $m$ is even,  according to \eref{C0-exp-1}, the tadpole coefficient of a tensor tadpole integral is given by
\bea
C_1^{(0)}(m)=c^{(m)}(M_0^2)^{{m/2}}s_{00}^{m/2}.
\eea
Since the tadpole coefficient only contains an unknown function $c^{(m)}$, only one recurrence relation is necessary. For a tensor tadpole integral, there is no external momenta $K_i$, so we can only consider the ${\cal T}$-type  recurrence relation  \eref{recurrence_relation_2}
\bea
4m (m-1) c^{(m-2)}=m(D+m-2)c^{(m)}, ~~~~~\label{tad-1}
\eea
then
\begin{align}
c^{(m)}= \frac{4 (m-1)}{(D+m-2)} c^{(m-2)}.
\end{align}
With the boundary condition $c^{(0)}=1$, we  immediately get
\bea
c^{(m)}=  {2^m (m-1)!!\over \prod_{i=1}^{m\over 2} (D+2(i-1))},\label{tad-coeff-2}\eea
so when $m$ is even, the tadpole coefficient is
\bea C_1^{(0)}(m)=c^{(m)}(M_0^2)^{m\over 2} s_{00}^{m\over 2}= M_0^{m} (R^2)^{m\over 2}{2^m (m-1)!!\over \prod_{i=1}^{m\over 2} (D+2(i-1))}.~~~\label{tad-coeff-3}\eea
This result \eref{tad-coeff-3} is consistent with that given by the traditional PV-reduction method in the Appendix \ref{sec:PVreduction}.

\subsection{Tadpole coefficients of tensor bubble integral}
%%%%%%%%%%%%%%%%%%%%%%
Now we consider the first nontrivial case, $n=1$ of \eref{Idea-1-1}, i.e., tensor bubble integral.
For now, the general form of tadpole coefficients \eref{C0-exp-1} becomes
\bea
C_1^{(0)}(m)=\sum^{\ \ \ \ \ \prime}_{i_0,i_1}c^{(m)}_{i_0,i_1}(M_0^2)^{i_0-1}s_{00}^{i_0}s_{01}^{i_1}=\sum_{i}c^{(m)}_{i}(M_0^2)^{{m-i\over 2}-1}s_{00}^{m-i\over 2}s_{01}^i,  \label{Bub_tadpole_expansion}
\eea
where in the second equation we have used the constraint $2i_0+i_1=m$ to solve $i_0$. Note that when $m$ is an even integer,  $i$ must be even, while if $m$ is an odd integer,  $i$ must be odd.
We consider the recurrence relations resulted by the differential operator $\mathcal{D}_i$.
Since $n=1$, \eref{recurrence_relation_1} can just give one recurrence relation
\bea
(i+1) \b_{11} c^{(m)}_{i+1}+ (m-i+1) c^{(m)}_{i-1}=m\a_1c^{(m-1)}_i-m\delta_{0,i}c^{(m-1)}.~~\label{Bub-D1-Relation}
\eea
We should note that $c^{(m-1)}$ in \eref{Bub-D1-Relation}  is the known expansion coefficients appearing in the reduction of tensor tadpole integrals (see \eref{tad-coeff-2}).
Replacing $i$ by $i+1$, we get
\bea
c^{(m)}_{i+2}={1\over (i+2)\beta_{11}}\left(m\a_1c^{(m-1)}_{i+1}-m\delta_{0,i+1}c^{(m-1)}-(m-i)c^{(m)}_i\right).~\label{Bub-ToNextTerm}
\eea
Note that $c^{(m)}_i$ is nonzero only for $0\le i \le m$ and $m-i$ is even.

For a certain rank $m$, by the \textit{inductive assumption} given in the subsection \ref{sec:differential_equation}, the expansion coefficients $c^{(m-1)}_{i+1}, c^{(m-1)}$ are already known in equation \eref{Bub-ToNextTerm},
then  equation \eref{Bub-ToNextTerm} has established the relation between $c^{(m)}_{i+2}$ and $c^{(m)}_{i}$. 
%relates expansion coefficient $c^{(m)}_{i+2}$ with $c^{(m)}_i$, in other words, the calculation of expansion coefficient $c^{(m)}_{i+2}$ is reduced into the calculation of $c^{(m)}_i$.
Then according to $m=2r$ or $m=2r+1$ with $r$ being a positive integer, the expansion coefficient $c^{(m)}_{i+2}$ is reduced into $c^{(2r)}_0$ or $c^{(2r+1)}_1$.
Now the task becomes the determination of the initial expansion coefficient $c^{(2r)}_0$ or $c^{(2r+1)}_1$.
For  $m=2r+1,i=0$, equation \eref{Bub-D1-Relation} becomes
\bea
c^{(2r+1)}_1={2r+1\over \beta_{11}}\left(\a_1c_0^{(2r)}-c^{(2r)}\right),~~~\label{Bub-Odd-FirstTerm}
\eea
which means that we can obtain $c_1^{(2r+1)}$ from $c_0^{(2r)}$ and $c^{(2r)}$. Thus we only need to calculate $c_0^{(2r)}$, and from $c_0^{(2r)}$ we can get all expansion coefficients by \eref{Bub-Odd-FirstTerm} and \eref{Bub-ToNextTerm}.

Now we consider the computation of $c_0^{(m)}$ with $m=2 r$. When $n=1$, the recurrence relation \eref{recurrence_relation_2} of the differential operator $\mathcal{T}$  with $m=2r, i=0$ becomes
\bea
r(D+2r-2
)c^{(2r)}_0+ \b_{11}c^{(2r)}_{2}=4r(2r-1)c_0^{(2r-2)}. \label{Bub-Even-FirstTerm-2}
\eea
Combining  \eref{Bub-ToNextTerm} with $i=0$, i.e.,
\begin{align}
c^{(2r)}_{2}
=&{r\over \beta_{11}}\left(\a_1c^{(2r-1)}_{1}-c^{(2r)}_0\right)
={r\over \beta_{11}}\left[\a_1{2r-1\over \beta_{11}}\left(\a_1c_0^{(2r-2)}-c^{(2r-2)}\right)-c^{(2r)}_0\right].~~~\label{Bub-Even-FirstTerm-1}
\end{align}
we can solve   $c^{(2r)}_0$
\begin{align}
c_0^{(2r)}
=\frac{2r-1}{2r+D-3}\left[ \left(4-\frac{\a_1^2}{\b_{11}} \right )c_0^{(2r-2)} + \frac{\a_1}{\b_{11}} c^{(2r-2)} \right], ~\label{Bub-Even-0Term}
\end{align}
with 
\bea \alpha_1\equiv f_1/M_0^2,~~~~~ \beta_{11}\equiv s_{11}/M_0^2\eea
With the obvious boundary condition $c^{(0)}_0=0$,  this recurrence relation \eref{Bub-Even-0Term} can be solved as
\begin{align}
c_0^{(2r)}= \frac{\a_1}{\b_{11}}\sum_{i=1}^r \left(\prod_{j=i}^r \frac{2j-1}{2j+D-3} \right) \left( 4-\frac{\a_1^2}{\b_{11}} \right )^{r-i}  c^{(2i-2)},  \label{Bub-Even-General-c0}
\end{align}
%

%
%\bea
%c_0^{(2r)}={(2r-1)!!\a_1\over(D+2r-3)\b_{11}}P^{(r)}(u)
%~~~~\label{Bub-Even-General-c0}\eea
%% % % % %
%where $u \equiv 4-\frac{\a_1^2}{\b_{11}}$ and  $P^{(r)}(u)\equiv \sum_{i=0}^{r-1}a^{(r)}_iu^i$ is a polynomial of $u$ with coefficients
%\bea
%a^{(r)}_i&=&{4^{r-i-1}  \over \prod_{k=0}^{r-2-i}(D+2k)\prod_{j=0}^{i-1}(D+2r-5-2j)}.
%\eea

%%%%%%%% figure
%%%%%%%%
\begin{figure}
\begin{center}
\begin{tikzpicture}[scale=1.3]
\newcommand{\rectangle}[4]{\draw[fill,black] (#1-0.5*#3,#2-0.5*#4)--(#1+0.5*#3,#2-0.5*#4)--(#1+0.5*#3,#2+0.5*#4)--(#1-0.5*#3,#2+0.5*#4)--cycle;}
\def\size{0.05}
\def\boxsize{0.07}
\def\offset{0.05}
\newcommand{\Di}[3]{\draw[->,line width=2,#3](#1-1+\offset,#2-1+\offset)to(#1-\offset,#2-\offset);
			\ifnum #2>1
			\draw[->,line width=2,cyan](#1+\offset,#2-2+\offset)[bend right]to(#1+\offset,#2-\offset);
			\fi}
%%%%%%
\draw[line width=1.5,->](-1,0)--(6,0);
\draw[line width=1.2,->](0,0)--(0,6);
			\draw[fill,green] (0,0) circle [radius=0.08];
		\foreach \x in {-1,0,1,2,...,5}{
				\node[below] at (\x,-0.1) {\x};
				\ifnum \x>0
				\node[left] at (-0.1,\x) {\x};
				\fi
				\node[below] at (\x,-0.1) {\x};
			\foreach \y in {-1,0,1,2,...,5}{
				\ifnum \y>\x
				\rectangle{\x}{\y}{\boxsize}{\boxsize};
				\fi
				\ifnum \y <0
				\rectangle{\x}{\y}{\boxsize}{\boxsize};
				\fi
			}

		}
	    \rectangle{1}{0}{\boxsize}{\boxsize};
	    \rectangle{2}{1}{\boxsize}{\boxsize};
	    \rectangle{3}{0}{\boxsize}{\boxsize};
	    \rectangle{3}{2}{\boxsize}{\boxsize};
	    \rectangle{4}{1}{\boxsize}{\boxsize};
	    \rectangle{4}{3}{\boxsize}{\boxsize};
	    \rectangle{5}{0}{\boxsize}{\boxsize};
	    \rectangle{5}{2}{\boxsize}{\boxsize};
	    \rectangle{5}{4}{\boxsize}{\boxsize};
	    \draw[fill,blue] (1,1) circle [radius=\size];
        \draw[fill,red] (2,0) circle [radius=\size];
        \draw[fill,blue] (2,2) circle [radius=\size];
        \draw[fill,blue] (3,1) circle [radius=\size];
        \draw[fill,blue] (3,3) circle [radius=\size];
        \draw[fill,red] (4,0) circle [radius=\size];
        \draw[fill,blue] (4,2) circle [radius=\size];
        \draw[fill,blue] (4,4) circle [radius=\size];
        \draw[fill,blue] (5,1) circle [radius=\size];
        \draw[fill,blue] (5,3) circle [radius=\size];
        \draw[fill,blue] (5,5) circle [radius=\size];
        \draw[->,line width=2,red](0+\offset,0-\offset)[bend right]to(2-\offset,0-\offset);
         \draw[->,line width=2,red](2+\offset,0-\offset)[bend right]to(4-\offset,0-\offset);
          \Di{1}{1}{orange};
           \Di{2}{2}{cyan};
         \Di{3}{1}{orange};
          \Di{3}{3}{cyan};
           \Di{4}{2}{cyan};
          \Di{4}{4}{cyan};
           \Di{5}{1}{orange};
           \Di{5}{3}{cyan};
           \Di{5}{5}{cyan};
\node[below] at (5.8,0) {$m$};
\node[left] at (0,5.8) {$i$};
\end{tikzpicture}

\caption{Algorithm for the calculation of the expansion coefficients of tadpole coefficients of tensor bubble  integral.
Here each point $(i,m)$ represents an expansion coefficient $c_{i}^{(m)}$.
The points represented by black squares are the zero expansion coefficients, while the points represented by blue or red circles are the unknown expansion coefficients we need to calculate.
The red thick arrow represents the recurrence relation \eref{Bub-Even-0Term}, the orange thick arrows represents the recurrence relation  \eref{Bub-Odd-FirstTerm}, and the cyan thick arrow represents the recurrence relation \eref{Bub-ToNextTerm}.}
\label{Bub-Figure}
\end{center}
\end{figure}
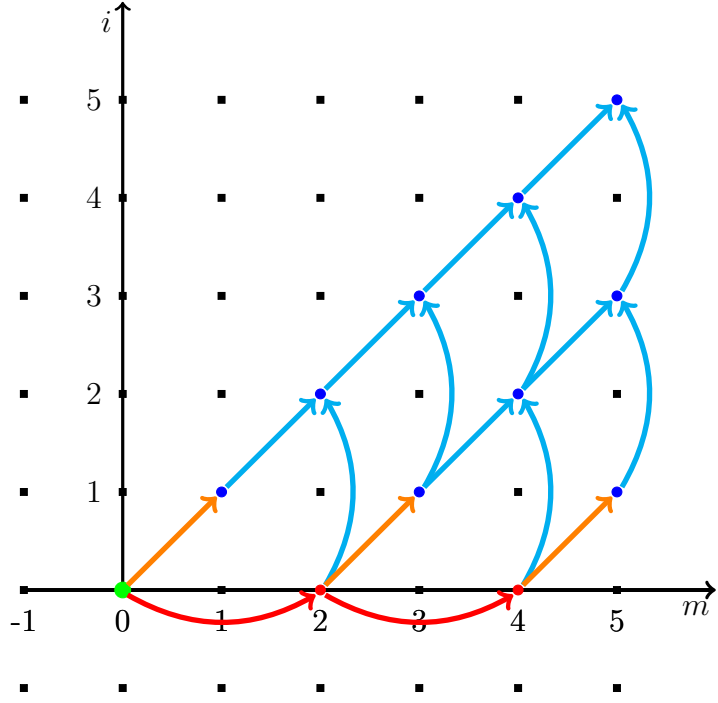
%%%%%%
%%%%%%

 After getting the analytic expression of $c^{(2r)}_0$,  we will show how to obtain other expansion coefficients $c_{i}^{(m)}$ by using the recurrence relation \eref{Bub-ToNextTerm} and \eref{Bub-Odd-FirstTerm} literately .
Let us take the expansion coefficients with rank from $1$ to $3$ as examples to illustrate the procedure of calculation.
\begin{itemize}

\item $m=1$: Only $c^{(1)}_1$ is necessary to calculate. Using \eref{Bub-Odd-FirstTerm}, we easily get
	\bea
	c^{(1)}_1={1\over \beta_{11}}\left(\a_1c_0^{(0)}-c^{(0)}\right)=-{1\over \beta_{11}}.
	\eea
	
\item $m=2$: $c^{(2)}_0$ and $c^{(2)}_2$ are unknown. First, $c^{(2)}_0$ is given directly by \eref{Bub-Even-0Term}
	\bea
	c_0^{(2)}&=&\frac{1}{D-1}\left[ \left(4-\frac{\a_1^2}{\b_{11}} \right )c_0^{(0)} + \frac{\a_1}{\b_{11}} c^{(0)} \right]={\a_1\over (D-1)\b_{11}}.
	\eea
then we use \eref{Bub-ToNextTerm} to calculate $c^{(2)}_2$
	\bea
	c^{(2)}_{2}={1\over 2\beta_{11}}\left(2\a_1c^{(1)}_{1}-2c^{(2)}_0\right)=-\frac{\alpha _1 D}{(D-1) \beta _{11}^2}.
	\eea
	
\item $m=3$: We need to calculate $c^{(3)}_1$ and $c^{(3)}_3$. First, we use \eref{Bub-Odd-FirstTerm} to calculate $c^{(3)}_1$
	\bea
	c^{(3)}_1={3\over \beta_{11}}\left(\a_1c_0^{(2)}-c^{(2)}\right)=\frac{3 \left(4 \beta _{11}-4 D \beta _{11}+\alpha _1^2 D\right)}{(D-1) D \beta _{11}^2},
	\eea
then we use \eref{Bub-ToNextTerm} to calculate $c^{(3)}_3$
	\bea
	c^{(3)}_{3}={1\over 3\beta_{11}}\left(3\a_1c^{(2)}_{2}-3c^{(3)}_1\right)=-\frac{8 \beta _{11}-8 D \beta _{11}+\alpha _1^2 D^2+2 \alpha _1^2 D}{(D-1) D \beta _{11}^3}.
	\eea
\end{itemize}
The above three examples are enough to illustrate the procedure of calculating expansion coefficients $c_i^{(m)}$.
Rough speaking, we calculate the expansion coefficients $c_i^{(m)}$ from lower rank to higher rank, and at a fixed rank $m$, we prefer to calculate the expansion coefficients $c_i^{(m)}$ with smaller index $i$ first.
More explicitly, our algorithm is summarized as following, as illustrate by fig.\ref{Bub-Figure}.

Supposing we want to calculate the tadpole coefficients with rank $m_0$, then we need to calculate all expansion coefficients with rank $m_0$,
\begin{itemize}

\item Step 1: Consider the rank $m=0$, which provides the boundary condition $c_0^{(0)}=0$.
	
\item Step 2: Consider the rank $m=1$, calculate the expansion coefficients $c^{(1)}_1$  by \eref{Bub-ToNextTerm} as illustrated before.

\item Step 3: Consider the rank $m=2$, calculate the expansion coefficients $c^{(2)}_0, c^{(2)}_2$ successively by \eref{Bub-Even-General-c0} and \eref{Bub-ToNextTerm} as illustrated before.

\item Step 4: Consider the rank $m=3$,  calculate the expansion coefficients $c^{(3)}_1, c^{(3)}_3$ successively by  \eref{Bub-ToNextTerm} as illustrated before.

 $\cdots$

\item Step $m \le m_0$:
	 If $m=2r$, calculate $c_{0}^{(2r)}, c_{2}^{(2r)},\cdots,c_{2r}^{(2r)}$ successively by using \eref{Bub-Even-General-c0} and \eref{Bub-ToNextTerm}.
	If $m=2r+1$, calculate $c_{0}^{(2r+1)}, c_{1}^{(2r+1)}, \cdots,c_{2r+1}^{(2r+1)}$ successively by using  \eref{Bub-ToNextTerm}.
	
    \item Final step: Combine all expansion coefficients to get the tadpole coefficient by \eref{Bub_tadpole_expansion}.
	
\end{itemize}

It's obvious that with  the help of Mathematica, one can easily implement the three key relations
\eref{Bub-Even-General-c0},  and \eref{Bub-ToNextTerm}, thus the analytic expression of tadpole coefficients can be nicely generated.

%to see that starting from the initial point, we can calculate all expansion coefficients following the above algorithm.
%With the help of Mathematica, one can easily implement the three key relations 
%\eref{Bub-Even-General-c0}, \eref{Bub-Odd-FirstTerm} and \eref{Bub-ToNextTerm}, thus
%the analytic expression of tadpole c

%%%%%%%%%%%%%%%%%%%%%
\subsection{Tadpole coefficients of tensor triangle integral}
%%%%%%%%%%%%%%%%%%%%%%
\label{sec:tensor triangle}
Now we consider the second nontrivial case, $n=2$ for \eref{Idea-1-1}, i.e., tensor triangle Feynman integral. In this case, the general form of the tadpole coefficients \eref{C0-exp-1} can be written as
\bea
C_1^{(0)}(m)=\sum_{i_1,i_2}(M_0^2)^{-2}  c_{i_1, i_2}^{(m)}(M_0^2s_{00})^{m-i_1-i_2\over 2}s_{01}^{i_1}s_{02}^{i_2}.  \label{Tri_tadpole_expansion}
\eea
As a result, the $\cal{D}$-type recurrence relations coming from ${\cal D}_i, i=1,2$ with  $n=2$ of \eref{recurrence_relation_1} are
\begin{align}
(m+1-i_1-i_2)c^{(m)}_{i_1-1,i_2} &+ (i_1+1) \b_{11}c^{(m)}_{i_1+1,i_2}+(i_2+1) \b_{12}c^{(m)}_{i_1,i_2+1} \nn
& =m\a_1c^{(m-1)}_{i_1,i_2}-m\delta_{i_1,0}c^{(m-1)}_{i_2}[ \hat{1}],\nn
(m+1-i_1-i_2)c^{(m)}_{i_1,i_2-1} &+ (i_1+1) \b_{12}c^{(m)}_{i_1+1,i_2}+(i_2+1) \b_{22}c^{(m)}_{i_1,i_2+1} \nn
& =m\a_2c^{(m-1)}_{i_1,i_2}-m\delta_{i_2,0}c^{(m-1)}_{i_1}[ \hat{2}], \label{Tri-D2-Relation}
\end{align}
where we have added $[\widehat{1}]$ or $[\widehat{2}]$ behind $c^{(m-1)}_{i_2}$ or $c^{(m-1)}_{i_1}$ to stress that  it corresponds to the tadpole coefficients of the integrals with $1$st or $2$nd propagator being canceled. The above formulas can be rewritten in a compact form as\footnote{When we meet the multiplication of such vectors and Gram matrix, we always consider it as the matrix multiplication by default.}
\begin{align}
G(1,2)\left(\begin{array}{l}
(i_1+1)c^{(m)}_{i_1+1,i_2} \\
(i_2+1)c^{(m)}_{i_1,i_2+1}\\
\end{array} \right) =     \boldsymbol{O}^{(m)}(i_1,i_2).  \label{Tri-Di-Matrix}
\end{align}
where we use the notation
\begin{align}
\left( \boldsymbol{O}^{(m)}(i_1,i_2)  \right)^T \equiv \left(  O_1^{(m)}(i_1,i_2),O_2^{(m)}(i_1,i_2)  \right)
\end{align}
with
\begin{align}
O_1^{(m)}(i_1,i_2)&=m\left [\a_1c^{(m-1)}_{i_1,i_2}-\delta_{0,i_1}c^{(m-1)}_{i_2}[\hat{1}]\right]-(m+1-i_1-i_2)c^{(m)}_{i_1-1,i_2},\nn
O_2^{(m)}(i_1,i_2)&=m\left [\a_2c^{(m-1)}_{i_1,i_2}-\delta_{0,i_2}c^{(m-1)}_{i_1}[\hat{2}]\right]-(m+1-i_1-i_2)c^{(m)}_{i_1,i_2-1}.
\end{align}
Here we have defined the rescaled Gram matrix $G(1,2,\cdots,n)$ with $(i,j)$ element being
\begin{align}
[G(1,2,\cdots,n)]_{ij} \equiv \beta_{ij}=s_{ij}/M_0^2, \quad 1\le i,j\le n.
\end{align}
And we will denote its corresponding determinant as $\Delta(1,2,\cdots,n)$ and its $(i,j)$ cofactor as $\Delta^{(n)}_{i,j}$.

Note that two equations of \eref{Tri-Di-Matrix} are symmetric for exchanging $(1,2)$, so its result is also symmetric for $(1,2)$.
We can easily solve $c^{(m)}_{i_1+1,i_2},c^{(m)}_{i_1,i_2+1}$ as
\begin{align}
\left(\begin{array}{l}
c^{(m)}_{i_1+1,i_2} \\
c^{(m)}_{i_1,i_2+1}\\
\end{array} \right) = \left(\begin{array}{ll}
{1\over i_1+1}&0\\
0&{1\over i_2+1}\\
\end{array} \right)   G^{-1}   \boldsymbol{O}^{(m)}(i_1,i_2) . \label{Tri-Next-Term-Matrix}
\end{align}
More explicitly, they are
\begin{align}
c^{(m)}_{i_1+1,i_2}
&={1\over(i_1+1)\Delta(1,2)}   \Big[   (m+1-i_1-i_2) \Big(\b_{12}c^{(m)}_{i_1,i_2-1}-\b_{22}c^{(m)}_{i_1-1,i_2}\Big)\nn
&+m \Big( \beta _{12} \delta _{0,i_2}c^{(m-1)}_{i_1}[ \hat{2}] -\beta _{22} \delta _{0,i_1}c^{(m-1)}_{i_2}[ \hat{1}]  \Big)
+m \Big(\alpha_1\b_{22}-\a_2\b_{12}\Big)c^{(m-1)}_{i_1,i_2 }  \Big]\nn
c^{(m)}_{i_1,i_2+1}
&={1\over(i_2+1)\Delta(1,2)}   \Big[  (m+1-i_1-i_2)  \Big(\b_{12}c^{(m)}_{i_1-1,i_2}-\b_{11}c^{(m)}_{i_1,i_2-1} \Big)\nn
&+m \Big(\beta _{12} \delta _{0,i_1}c^{(m-1)}_{i_2}[\hat{1}] -\beta _{11} \delta _{0,i_2}c^{(m-1)}_{i_1}[\hat{2}]  \Big)
+m \Big(\alpha_2\b_{11}-\a_1\b_{12} \Big)c^{(m-1)}_{i_1,i_2} \Big]. \label{Tri-ToNextTerm}
\end{align}

For a certain rank $m$, by the \textit{inductive assumption} given in the subsection \ref{sec:differential_equation}, the expansion coefficients $c^{(m-1)}_{i_1,i_2}, c^{(m-1)}_{i_1}$ and $c^{(m-1)}_{i_2}$ are already known in equation \eref{Tri-ToNextTerm}, hence equation \eref{Tri-ToNextTerm} relates expansion coefficient $c^{(m)}_{i_1+1,i_2}$, $c^{(m)}_{i_1,i_2+1}$ to $c^{(m)}_{i_1-1,i_2}$, $c^{(m)}_{i_1,i_2-1}$. 
%In other words, the calculation of expansion coefficients $c^{(m)}_{i_1+1,i_2}$, $c^{(m)}_{i_1,i_2+1}$ is reduced into the calculation of $c^{(m)}_{i_1-1,i_2}$, $c^{(m)}_{i_1,i_2-1}$ with smaller indices.
So according to rank $m$ is even or odd, the expansion coefficient $c^{(m)}_{i_1,i_2}$ will be iteratively reduced into the algebraic expressions of $c^{(m)}_{0,0}$ or $c^{(m)}_{1,0}$ and $c^{(m)}_{0,1}$ finally. In the odd rank case, $c^{(m)}_{0,1}$ can be got from $c^{(m)}_{1,0}$ by using the previously mentioned symmetry.
Now the task becomes the determination of the initial expansion coefficient $c^{(m)}_{0,0}$ for $m=2r$  and $c^{(m)}_{1,0}$ for $m=2r+1$.

First, we consider the odd rank case with $m=2r+1$. When $i_1=i_2=0$, equation \eref{Tri-ToNextTerm} becomes
\bea
c^{(2r+1)}_{1,0}&=&{2r+1\over\Delta(1,2)} \Big[  \beta _{12}c^{(2r)}_0[\hat{2}] -\beta _{22} c^{(2r)}_0[\hat{1}]
+(\alpha_1\b_{22}-\a_2\b_{12})c^{(2r)}_{0,0}\Big],\nn
c^{(2r+1)}_{0,1}&=&{2r+1\over\Delta(1,2)} \Big[  \beta _{12} c^{(2r)}_0[\hat{1}] -\beta _{11} c^{(2r)}_0[\hat{2}]
+(\alpha_2\b_{11}-\a_1\b_{12})c^{(2r)}_{0,0}  \Big],~~~~\label{Tri-Odd-FirstTerm}
\eea
which means that we can obtain $c_{1,0}^{(2r+1)}$ and $c_{0,1}^{(2r+1)}$  from $c_{0,0}^{(2r)}$ and $c^{(2r)}_0$. So we only need to calculate $c_{0,0}^{(2r)}$, and from it we can get all expansion coefficients by \eref{Tri-Odd-FirstTerm} and \eref{Tri-ToNextTerm} recursively.

Now we consider the computation of $c_{0,0}^{(2r)}$. To calculate it, we need the $\cal T$-type recurrence relations 
 \begin{align}
8r(2r-1)c^{(2r-2)}_{i_1,i_2}
=& (2r-i_1-i_2)(D+2r+i_1+i_2-2)c^{(2r)}_{i_1,i_2} +2(i_1+1)(i_2+1)\b_{12}c^{(2r)}_{i_{1}+1,i_2+1} \notag\\
&+ (i_1+1)(i_1+2)\b_{11}c^{(2r)}_{i_1+2,i_2}+ (i_2+1)(i_2+2)\b_{22}c^{(2r)}_{i_1,i_2+2}. \label{Tri_recurrence_relation_2}
\end{align}
When $i_1=i_2=0$, the recurrence relation \eref{Tri_recurrence_relation_2} becomes
\bea
r(2r+D-2)c^{(2r)}_{0,0}+\b_{11}c^{(2r)}_{2,0}+\b_{22}c^{(2r)}_{0,2}+\b_{12}c^{(2r)}_{1,1}=4r(2r-1)c^{(2r-2)}_{0,0}.~~~\label{Tri-T-c00}
\eea
Since the above equation contains three unknown coefficients $c_{0,2}^{(2r)}$ , $c_{2,0}^{(2r)}$ and $c_{1,1}^{(2r)}$ with larger indices, we need to reduce them to expansion coefficients with smaller indices. Using \eref{Tri-ToNextTerm}  we have
\bea
c^{(2r)}_{2,0}&=&{r\over\Delta(1,2)} \Big[ -\b_{22}c^{(2r)}_{0,0}+ \beta _{12} c^{(2r-1)}_{1}[\hat{2}] +(\alpha_1\b_{22}-\a_2\b_{12})c^{(2r-1)}_{1,0}\Big],\nn
c^{(2r)}_{0,2}&=&{r\over\Delta(1,2)} \Big[-\b_{11}c^{(2r)}_{0,0}+ \beta _{12} c^{(2r-1)}_{1}[\hat{1}] +(\alpha_2\b_{11}-\a_1\b_{12})c^{(2r-1)}_{0,1}\Big],\nn
c^{(2r)}_{1,1}&=&{2r\over\Delta(1,2)} \Big[\b_{12}c^{(2r)}_{0,0} -\beta _{22} c^{(2r-1)}_1[\hat{1}]+(\alpha_1\b_{22}-\a_2\b_{12})c^{(2r-1)}_{0,1}\Big],
\eea
then we use \eref{Bub-ToNextTerm} and \eref{Tri-Odd-FirstTerm} to reduce the odd rank $(2r-1)$ expansion coefficients to $2r-2$ rank expansion coefficients. After combining the results and \eref{Tri-T-c00}, we finally get a recurrence relation of  $c^{(2r)}_{0,0}$ with respect to $r$ as
\begin{align}
c^{(2r)}_{0,0}
=&\frac{2 r-1}{2r+D-4} \left[ \frac{2 \a_1 \a_2 \b_{12}-\a_2^2 \b_{11}-\a_1^2 \b_{22} }{\Delta(1,2) }+4  \right]
c^{(2r-2)}_{0,0}\nn
&+\frac{2 r-1}{(2r+D-4) \Delta(1,2)  }  \left[ \left(\a_2 \b_{11}-\a_1 \b_{12}\right) c^{(2r-2)}_0[\hat{2}]+\left(\a_1 \b_{22}-\a_2 \b_{12}\right)c^{(2r-2)}_0[\hat{1}] \right] \notag\\
=& \frac{2 r-1}{2r+D-4} \left[(4- \boldsymbol{\alpha}^T  G^{-1} \boldsymbol{\alpha}) c^{(2r-2)}_{0,0}   +   \boldsymbol{\alpha}^T  G^{-1} \boldsymbol{c}_0^{(2r-2)} \right],~~~\label{Tri-Even-FirstTerm}
\end{align}
where $\boldsymbol{\alpha}^T=(\alpha_1,\alpha_2)$ and $(\boldsymbol{c}_0^{(2r-2)})^T= (c^{(2r-2)}_0[\hat{1}], c^{(2r-2)}_0[\hat{2}])$ are two vectors, and $G^{-1}$ is the inverse of Gram matrix $G(1,2)$. Using the boundary condition $c_{0,0}^{(0)}=0$, it's easy to get the expression for $c^{(2r)}_{0,0}$ as
\begin{align}
c^{(2r)}_{0,0}= \sum_{i=2}^r \left(\prod_{j=i}^r \frac{2j-1}{2j+D-4}\right) \left(4- \boldsymbol{\alpha}^T  G^{-1} \boldsymbol{\alpha} \right)^{r-i}  \left(\boldsymbol{\alpha}^T  G^{-1} \boldsymbol{c}_0^{(2i-2)} \right),    \label{Tri_expression_even}
\end{align}
where we have used the boundary condition $(\boldsymbol{c}_0^{(0)})^T=(0,0)$, so the summation is from $i=2$ to $r$.

%%%%%%%%%% algorithm for triangle
\begin{figure}
	\centering
        \subfigure[$m=5$]
        { %
        \begin{tikzpicture}[scale=0.7]
        \newcommand{\rectangle}[4]{
        	\draw[fill,black] (#1-0.5*#3,#2-0.5*#4)--(#1+0.5*#3,#2-0.5*#4)--(#1+0.5*#3,#2+0.5*#4)--(#1-0.5*#3,#2+0.5*#4)--cycle;
        }
        \def\size{0.05}
        \def\boxsize{0.07}
        \def\offset{0.1}
        \newcommand{\Difirst}[2]{
        	\ifnum #1>0
        	\ifnum #2>0
        	\draw[->,line width=2,cyan](#1-1+\offset,#2-1+\offset)to(#1-\offset,#2-\offset);
        	\fi
        	\fi
        	\ifnum #2>1
        	\draw[->,line width=2,cyan](#1,#2-2+\offset)to(#1,#2-\offset);
        	\fi
        }
        \newcommand{\Disecond}[2]{
        	\ifnum #1>0
        	\ifnum #2>0
        	\draw[->,line width=2,red](#1-1+\offset,#2-1+\offset)to(#1-\offset,#2-\offset);
        	\fi
        	\fi
        	\ifnum #1>1
        	\draw[->,line width=2,red](#1-2+\offset,#2)to(#1-\offset,#2);
        	\fi
        }
        \newcommand{\upright}[2]{
        	\draw[->,thick,red] (#1+0.1,#2+0.1)--(#1+0.9,#2+0.9);
        }
        \draw[->,thick](-1.6,0)--(6.7,0);
        \draw[->,thick](0,-1.6)--(0,6.7);
        \foreach \t in {1,3,...,5}{
        	\node[below] at (\t,-0.2) {\t};
        	
        }
        \foreach \t in {1,3,...,5}{
        	\node[left] at (-0.1,\t) {\t};
        }
        \draw[fill,green] (1,0) circle [radius=0.10];
        \draw[fill,green] (0,1) circle [radius=0.10];
        \draw[fill,blue] (3,0) circle [radius=0.07];
        \draw[fill,blue] (0,3) circle [radius=0.07];
        \draw[fill,blue] (1,2) circle [radius=0.07];
        \draw[fill,blue] (2,1) circle [radius=0.07];
        \draw[fill,blue] (5,0) circle [radius=0.07];
        \draw[fill,blue] (0,5) circle [radius=0.07];
        \draw[fill,blue] (1,4) circle [radius=0.07];
        \draw[fill,blue] (2,3) circle [radius=0.07];
        \draw[fill,blue] (3,2) circle [radius=0.07];
        \draw[fill,blue] (4,1) circle [radius=0.07];
        \def\size{0.10};
        \rectangle{-2}{-1}{\size}{\size};
        \rectangle{0}{-1}{\size}{\size};
        \rectangle{2}{-1}{\size}{\size};
        \rectangle{4}{-1}{\size}{\size};
        \rectangle{6}{-1}{\size}{\size};
        \rectangle{-1}{0}{\size}{\size};
        \rectangle{6}{1}{\size}{\size};
        \rectangle{-2}{1}{\size}{\size};
        \rectangle{-1}{2}{\size}{\size};
        \rectangle{5}{2}{\size}{\size};
        \rectangle{4}{3}{\size}{\size};
        \rectangle{6}{3}{\size}{\size};
        \rectangle{3}{4}{\size}{\size};
        \rectangle{5}{4}{\size}{\size};
        \rectangle{4}{5}{\size}{\size};
        \rectangle{6}{5}{\size}{\size};
        \rectangle{2}{5}{\size}{\size};
        \rectangle{1}{6}{\size}{\size};
        \rectangle{3}{6}{\size}{\size};
        \rectangle{5}{6}{\size}{\size};
        \rectangle{-1}{4}{\size}{\size};
        \rectangle{-1}{6}{\size}{\size};
        \rectangle{-2}{3}{\size}{\size};
        \rectangle{-2}{5}{\size}{\size};
        \draw[dashed](1,0)--(0,1);
        \draw[dashed](3,0)--(0,3);
        \draw[dashed](5,0)--(0,5);

        \Difirst{0}{3};
        \Difirst{0}{5};
        \Difirst{1}{2};
        \Difirst{2}{1};
        \Difirst{0}{5};
        \Difirst{2}{3};
        \Difirst{3}{2};
        \Difirst{4}{1};
        \Difirst{1}{4};
        \Disecond{1}{0};
        \Disecond{3}{0};
        \Disecond{5}{0};
        \node[below] at (6.8,0) {$i$};
        \node[left] at (0,6.8) {$j$};
        \end{tikzpicture}
    }
        	\subfigure[$m=6$]
        {%
        	\begin{tikzpicture}[scale=0.7]
        	\newcommand{\rectangle}[4]{
        		\draw[fill,black] (#1-0.5*#3,#2-0.5*#4)--(#1+0.5*#3,#2-0.5*#4)--(#1+0.5*#3,#2+0.5*#4)--(#1-0.5*#3,#2+0.5*#4)--cycle;
        	}
        	\def\size{0.05}
        	\def\boxsize{0.07}
        	\def\offset{0.1}
        	\newcommand{\Difirst}[2]{
        		\ifnum #1>0
        		\ifnum #2>0
        		\draw[->,line width=2,cyan](#1-1+\offset,#2-1+\offset)to(#1-\offset,#2-\offset);
        		\fi
        		\fi
        		\ifnum #2>1
        		\draw[->,line width=2,cyan](#1,#2-2+\offset)to(#1,#2-\offset);
        		\fi
        	}
        	\newcommand{\Disecond}[2]{
        		\ifnum #1>0
        		\ifnum #2>0
        		\draw[->,line width=2,red](#1-1+\offset,#2-1+\offset)to(#1-\offset,#2-\offset);
        		\fi
        		\fi
        		\ifnum #1>1
        		\draw[->,line width=2,red](#1-2+\offset,#2)to(#1-\offset,#2);
        		\fi
        	}
        	\newcommand{\upright}[2]{
        		\draw[->,thick,red] (#1+0.1,#2+0.1)--(#1+0.9,#2+0.9);
        	}
        	\draw[->,thick](-1.6,0)--(6.7,0);
        	\draw[->,thick](0,-1.6)--(0,6.7);
        	\draw[fill,green] (0,0) circle [radius=0.1];
        	\foreach \t in {0,2,4,...,6}{
        		\ifnum \t >0
        		\draw[fill,blue] (\t,0) circle [radius=0.07];
        		\fi
        		\node[below] at (\t,-0.2) {\t};
        		
        	}
        	\foreach \t in {2,4,...,6}{
        		\node[left] at (-0.1,\t) {\t};
        		\draw[fill,blue] (0,\t) circle [radius=0.07];
        	}
        	\draw[fill,blue] (1,1) circle [radius=0.07];
        	\draw[fill,blue] (1,3) circle [radius=0.07];
        	\draw[fill,blue] (3,1) circle [radius=0.07];
        	\draw[fill,blue] (2,2) circle [radius=0.07];
        	\draw[fill,blue] (1,5) circle [radius=0.07];
        	\draw[fill,blue] (5,1) circle [radius=0.07];
        	\draw[fill,blue] (3,3) circle [radius=0.07];
        	\draw[fill,blue] (2,4) circle [radius=0.07];
        	\draw[fill,blue] (4,2) circle [radius=0.07];
        	\foreach \t in {-2,0,2,4,...,6}{
        		%\rectangle{\t}{-1}{\size}{\size};
        		%\rectangle{\t}{1}{\size}{\size};
        		%\rectangle{\t}{3}{\size}{\size};
        		%\rectangle{\t}{5}{\size}{\size};
        		
        	}
        	\foreach \t in {0,2,4,...,6}{
        		%\rectangle{\t-1}{0}{\size}{\size};
        		%\rectangle{\t-1}{2}{\size}{\size};
        		%\rectangle{\t-1}{4}{\size}{\size};
        		%\rectangle{\t-1}{6}{\size}{\size};
        		
        	}
        	\def\size{0.10};
        	\rectangle{6}{2}{\size}{\size};
        	%\rectangle{6}{3}{\size}{\size};
        	\rectangle{6}{4}{\size}{\size};
        	\rectangle{6}{6}{\size}{\size};
        	\rectangle{5}{5}{\size}{\size};
        	\rectangle{5}{3}{\size}{\size};
        	%\rectangle{5}{4}{\size}{\size};
        	%\rectangle{5}{6}{\size}{\size};
        	\rectangle{4}{4}{\size}{\size};
        	\rectangle{4}{6}{\size}{\size};
        	\rectangle{3}{5}{\size}{\size};
        	\rectangle{2}{6}{\size}{\size};
        	\rectangle{5}{-1}{\size}{\size};
        	\rectangle{3}{-1}{\size}{\size};
        	\rectangle{1}{-1}{\size}{\size};
        	\rectangle{-1}{-1}{\size}{\size};
        	\rectangle{-1}{1}{\size}{\size};
        	\rectangle{-1}{3}{\size}{\size};
        	\rectangle{-1}{5}{\size}{\size};
        	\rectangle{-2}{0}{\size}{\size};
        	\rectangle{-2}{2}{\size}{\size};
        	\rectangle{-2}{4}{\size}{\size};
        	\rectangle{-2}{6}{\size}{\size};
        	\draw[dashed](2,0)--(0,2);
        	\draw[dashed](4,0)--(0,4);
        	\draw[dashed](6,0)--(0,6);
        	\upright{0}{0};
        	\Difirst{0}{2};
        	\Difirst{0}{4};
        	\Difirst{0}{6};
        	\Difirst{1}{1};
        	\Difirst{2}{2};
        	\Difirst{1}{3};
        	\Difirst{3}{1};
        	\Difirst{1}{5};
        	\Difirst{2}{4};
        	\Difirst{3}{3};
        	\Difirst{4}{2};
        	\Difirst{5}{1};
        	\Disecond{2}{0};
        	\Disecond{4}{0};
        	\Disecond{6}{0};
        	\node[below] at (6.8,0) {$i$};
        	\node[left] at (0,6.8) {$j$};
        	\end{tikzpicture}
        }
		\caption{Algorithm for the calculation of the expansion coefficients of tadpole coefficients of a tensor triangle  integral with rank $m=5$ and $m=6$. Here the each point $(i,j)$ represents an expansion coefficient $c_{i,j}^{(5)}$ and $c_{i,j}^{(6)}$. The points represented by black squares are the zero expansion coefficients (we don't draw those expansion coefficients with $i+j=even$ for $m=5$ and $i+j=odd$ for $m=6$, for they all vanish according to the definition) while the points represented by blue or green circles are the unknown expansion coefficients  we need to calculate. The  red thick arrow represents the first one of the recurrence relations  \eref{Tri-ToNextTerm} and the cyan thick arrow represents the second one of the recurrence relations \eref{Tri-ToNextTerm}.}
		\label{Tri-Figure}
	
\end{figure}
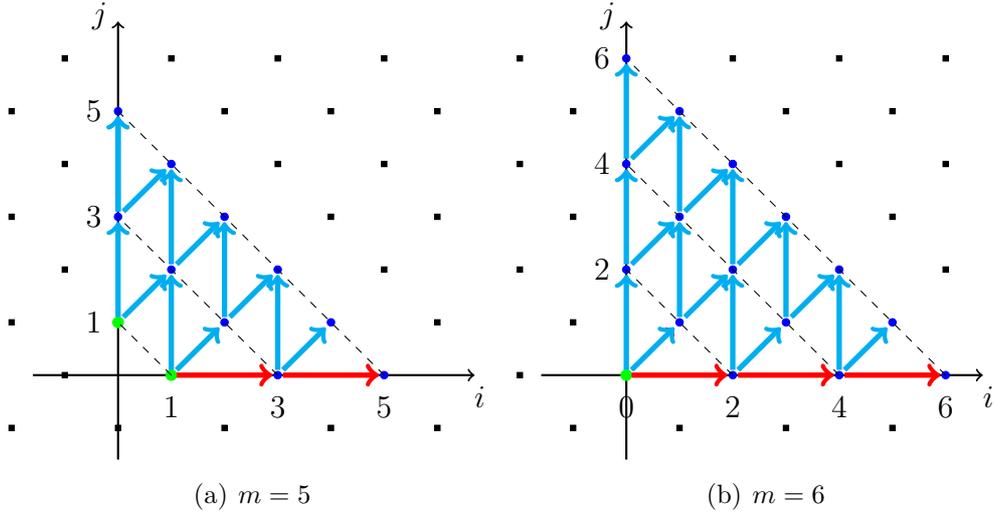
%%%%%%%%%%%%

%%%%%%%%%%
 After obtaining the analytic expression of $c_{0,0}^{(2r)}$ recursively using \eref{Tri-Even-FirstTerm},  we will consider how to obtain all expansion coefficients $c_{i,j}^{(m)}$ by using the recurrence relations \eref{Tri-ToNextTerm}, \eref{Tri-Odd-FirstTerm} and \eref{Tri-Even-FirstTerm} step by step.
First, we note that two equations of \eref{Tri-Odd-FirstTerm} can be rewritten as
\begin{align}
\left(\begin{array}{l}
c^{(2r+1)}_{1,0} \\
c^{(2r+1)}_{0,1}
\end{array} \right) = (2r+1)   G^{-1} \left( c_{0,0}^{(2r)}~  \boldsymbol{\alpha}  -  \boldsymbol{c}_0^{(2r)} \right),   \label{Tri_expression_odd}
\end{align}
since the expressions of  $c_{0,0}^{(2r)}$ and $\boldsymbol{c}_0^{(2r)}$ are known, then the results of $c^{(2r+1)}_{1,0}, c^{(2r+1)}_{0,1}$ are easy to get.
Second, we take several examples to illustrate the procedure of calculation.
\begin{itemize}
	\item $m=1$: We only need to calculate $c^{(1)}_{0,1}$ and $c^{(1)}_{1,0}$. Using \eref{Tri-Odd-FirstTerm} and $c^{(0)}_{0,0}=c^{(0)}_0[\hat{2}]=c^{(0)}_0[\hat{1}]=0$, it's trivial to get $c^{(1)}_{1,0}=c^{(1)}_{0,1}=0$.

	\item $m=2$: We need to calculate $c^{(2)}_{0,0},c^{(2)}_{1,1},c^{(2)}_{2,0},c^{(2)}_{0,2}$. From \eref{Tri_expression_even}, it's trivial to see $c^{(2)}_{0,0}=0$,
    then using \eref{Tri-ToNextTerm} and the symmetry we get
	\begin{align}
    c^{(2)}_{0,2}&={1\over2\Delta(1,2)} \Big(-2\b_{11}c^{(2)}_{0,0}   +2 \beta _{12} c^{(1)}_1[\hat{1}]  +2(\alpha_2\b_{11}-\a_1\b_{12})c^{(1)}_{0,1}\Big)
    =-\frac{\beta _{12}}{\beta _{22} \Delta (1,2)}, \nn
	c^{(2)}_{2,0}&=\left.c^{(2)}_{0,2}\right\vert_{1\leftrightarrow 2}=-\frac{\beta _{12}}{\beta _{11} \Delta (1,2)}, \nn
	c^{(2)}_{1,1}&={1\over\Delta(1,2)} \Big(2\b_{12}c^{(2)}_{0,0}-2\beta _{22} c^{(1)}_1[\hat{1}]
	+2(\alpha_1\b_{22}-\a_2\b_{12})c^{(1)}_{0,1}\Big)
	=\frac{2}{\Delta (1,2)}.
	\end{align}
	
	\item $m=3$: We need to calculate $c^{(3)}_{1,0},c^{(3)}_{0,1},c^{(3)}_{1,2},c^{(3)}_{2,1},c^{(2)}_{0,3},c^{(3)}_{3,0}$.
	First, we use \eref{Tri_expression_odd} to get
	\begin{align}
	\left(\begin{array}{l}
	c^{(3)}_{1,0} \\
	c^{(3)}_{0,1}
	\end{array} \right) = -3   G^{-1}   \boldsymbol{c}_0^{(2)}
	=\frac{ 3}{(D-1)\Delta (1,2) }\left(\begin{array}{l}
	\frac{ \alpha _1 \beta _{12}- \alpha _2 \beta _{11}}{ \beta _{11} } \\
	\frac{ \alpha _2 \beta _{12}- \alpha _1 \beta _{22}}{ \beta _{22} }
	\end{array} \right).
	\end{align}
	Next using \eref{Tri-ToNextTerm}, we have
	\begin{align}
	c^{(3)}_{1,2}&={1\over2\Delta(1,2)} \Big(2(\b_{12}c^{(3)}_{0,1}-\b_{11}c^{(3)}_{1,0})
	+3(\alpha_2\b_{11}-\a_1\b_{12})c^{(2)}_{1,1}\Big)\nn
	&=\frac{3 \alpha _2 D}{(D-1) \beta _{22} \Delta (1,2)}+\frac{3 (D+1) \beta _{12} \left(\alpha _2 \beta _{12}-\alpha _1 \beta _{22}\right)}{(D-1) \beta _{22} \Delta (1,2)^2},\nn
	c^{(3)}_{3,0}&={1\over3\Delta(1,2)} \Big(-2\b_{22}c^{(3)}_{1,0}
	+3 \beta _{12} c^{(2)}_2[\hat{2}]
	+3(\alpha_1\b_{22}-\a_2\b_{12})c^{(2)}_{2,0}\Big)\nn
	&=\frac{\alpha _2 \left(2 \beta _{11} \beta _{22}+(D-1) \beta _{12}^2\right)-\alpha _1 (D+1) \beta _{12} \beta _{22}}{(D-1) \beta _{11} \Delta (1,2)^2}-\frac{\alpha _1 D \beta _{12}}{(D-1) \beta _{11}^2 \Delta (1,2)}.
	\end{align}
	Other expansion coefficients can be got by using the permutation symmetry.
\end{itemize}
The above three examples are enough to illustrate the procedure of calculating expansion coefficients $c_{i,j}^{(m)}$.
Rough speaking, we calculate the expansion coefficients $c_{i,j}^{(m)}$ from lower rank to higher rank, and at a fixed rank $m$, we prefer to calculate the expansion coefficients $c_{i,j}^{(m)}$ with smaller $i+j$ first.
For a general rank $m_0$, we need to calculate all expansion coefficients with rank $m\le m_0$ to calculate the tadpole coefficient with rank $m_0$.
Now our algorithm of calculating tadpole coefficient with rank $m_0$ is summarized as following, as illustrate by fig.\ref{Tri-Figure}.
\begin{itemize}
	
	\item Step 1: Consider the rank $m=0$, which provides the boundary condition $c_{0,0}^{(0)}=0$.
	
	\item Step 2: Consider the rank $m=1$, the expansion coefficients $c^{(1)}_{0,1}, c^{(1)}_{1,0}$  are given by  \eref{Tri_expression_odd} as illustrated before.
	
	\item Step 3: Consider the rank $m=2$,  calculate the expansion coefficients $c^{(2)}_{0,0}, c^{(2)}_{1,1},c^{(2)}_{2,0},c^{(2)}_{0,2}$ successively by \eref{Tri-Even-FirstTerm} and \eref{Tri-ToNextTerm} as illustrated before.
	
	$\cdots$
	
	\item Step $m \le m_0$:
	If $m=2r$, calculate $c_{0,0}^{(2r)}, c_{1,1}^{(2r)},c_{2,0}^{(2r)},\cdots,c_{2r,0}^{(2r)}$ successively by using \eref{Tri-Even-FirstTerm} and \eref{Tri-ToNextTerm}.
	If $m=2r+1$, the equation directly gives the expressions of $c_{1,0}^{(2r+1)}, c_{0,1}^{(2r+1)}$, then calculate other expansion coefficients $c_{i,j}^{(2r+1)}$ successively by using \eref{Tri-ToNextTerm}.
	
	\item Final step: combine all expansion coefficients to get the tadploe coefficient by \eref{Tri_tadpole_expansion}.
	
\end{itemize}
With the help of Mathematica, one can easily implement the relations \eref{Tri-Even-FirstTerm} and \eref{Box-Next-Term-Matrix}, thus analytic reduction coefficients can be obtained for any rank. 

%such process can be easily done automatically.
%We can find that the algorithm is almost the same as the previous one in last subsection, which is not accident but and generally true for tensor integrals with more external momenta.

% % % % % % % % % % % % % %
\subsection{Tadpole coefficients of tensor box integral}
% % % % % % % % % % % % % %
Now we consider the third nontrivial case, $n=3$ of \eref{Idea-1-1}, i.e., the  tensor box Feynman integral, which is similar to the previous cases. In this case, the general form of the tadpole coefficients \eref{C0-exp-1} can be written as
\bea
C_1^{(0)}(m)=\sum_{i_1,i_2,i_3}(M_0^2)^{-3}(M_0^2s_{00})^{m-i_1-i_2-i_3\over 2}c^{(m)}_{i_1i_2i_3}s_{01}^{i_1}s_{02}^{i_2}s_{03}^{i_3}.\label{Box-Exp}
\eea
As a result, the $\cal{D}$-type recurrence relations are given by  setting $n=3$ in \eref{recurrence_relation_1}:
\begin{align}
(m+1-i_1-i_2-i_3)c^{(m)}_{i_1-1,i_2,i_3}&+(i_1+1)\b_{11}c^{(m)}_{i_1+1,i_2,i_3}+(i_2+1)\b_{12}c^{(m)}_{i_1,i_2+1,i_3} \nn
&+(i_3+1)\b_{13}c^{(m)}_{i_1,i_2,i_3+1}= m\left(\a_1c^{(m-1)}_{i_1i_2i_3}-\delta_{0,i_1}c^{(m-1)}_{i_2i_3}[\hat{1}]\right),\nn
(m+1-i_1-i_2-i_3)c^{(m)}_{i_1,i_2-1,i_3}
&+(i_1+1)\b_{12}c^{(m)}_{i_1+1,i_2,i_3}+(i_2+1)\b_{22}c^{(m)}_{i_1,i_2+1,i_3}\nn
&+(i_3+1)\b_{23}c^{(m)}_{i_1,i_2,i_3+1} =m\left(\a_2c^{(m-1)}_{i_1i_2i_3}-\delta_{0,i_2}c^{(m-1)}_{i_1i_3}[\hat{2}]\right),\nn
(m+1-i_1-i_2-i_3)c^{(m)}_{i_1,i_2,i_3-1}
&+(i_1+1)\b_{13}c^{(m)}_{i_1+1,i_2,i_3}+(i_2+1)\b_{23}c^{(m)}_{i_1,i_2+1,i_3}\nn
&+(i_3+1)\b_{33}c^{(m)}_{i_1,i_2,i_3+1}
=m\left(\a_3c^{(m-1)}_{i_1i_2i_3}-\delta_{0,i_3}c^{(m-1)}_{i_1i_2}[\hat{3}]\right),\no
\end{align}
where the notation $[\hat{i}]$ corresponds to the integral with $i$-th propagator being canceled. 
%It's easy to see that the three equations are symmetric for $\{1,2,3\}$, so the solutions of them also have the symmetry. 
These equations can be rewritten in a more compact form as
\begin{align}
	G\left(\begin{array}{l}
		(i_1+1)c^{(m)}_{i_1+1,i_2,i_3} \\
		(i_2+1)c^{(m)}_{i_1,i_2+1,i_3}\\
		(i_3+1)c^{(m)}_{i_1,i_2,i_3+1}\\
	\end{array} \right) =     \boldsymbol{O}^{(m)}(i_1,i_2,i_3) ,   \label{Box-Di-Matrix}
\end{align}
where we have used the notation $G\equiv (\beta_{ij})$ with $i,j=1,\cdots,3$ and defined the vector
\begin{align}
\left( \boldsymbol{O}^{(m)}(i_1,i_2,i_3)  \right)^T=\left(  O_1^{(m)}(i_1,i_2,i_3),O_2^{(m)}(i_1,i_2,i_3),O_3^{(m)}(i_1,i_2,i_3)  \right)
\end{align}
with
\begin{align}
O_1^{(m)}(i_1,i_2,i_3)&=m\left [\a_1c^{(m-1)}_{i_1,i_2,i_3}-\delta_{0,i_1}c^{(m-1)}_{i_2,i_3}[\hat{1}]\right]-(m+1-i_1-i_2-i_3)c^{(m)}_{i_1-1,i_2,i_3},\nn
O_2^{(m)}(i_1,i_2,i_3)&=m\left [\a_2c^{(m-1)}_{i_1,i_2,i_3}-\delta_{0,i_2}c^{(m-1)}_{i_1,i_3}[\hat{2}]\right]-(m+1-i_1-i_2-i_3)c^{(m)}_{i_1,i_2-1,i_3},\nn
O_3^{(m)}(i_1,i_2,i_3)&=m\left [\a_3c^{(m-1)}_{i_1,i_2,i_3}-\delta_{0,i_3}c^{(m-1)}_{i_1,i_2}[\hat{3}]\right]-(m+1-i_1-i_2-i_3)c^{(m)}_{i_1,i_2,i_3-1}.
\end{align}
Then we can easily solve $c^{(m)}_{i_1+1,i_2,i_3},c^{(m)}_{i_1,i_2+1,i_3}$ and $c^{(m)}_{i_1,i_2,i_3+1}$ by \eref{Box-Di-Matrix}  as
\begin{align}
	\left(\begin{array}{l}
		c^{(m)}_{i_1+1,i_2,i_3} \\
		c^{(m)}_{i_1,i_2+1,i_3}\\
		c^{(m)}_{i_1,i_2,i_3+1}\\
	\end{array} \right) = \left(\begin{array}{lll}
		{1\over i_1+1}&0&0\\
		0&{1\over i_2+1}&0\\
		0&0&{1\over i_3+1}\\
	\end{array} \right)   G^{-1}   \boldsymbol{O}^{(m)}(i_1,i_2,i_3) . \label{Box-Next-Term-Matrix}
\end{align}
More explicitly, they are
\begin{align}
	c^{(m)}_{i_1+1,i_2,i_3}&={1\over (i_1+1)\Delta(1,2,3)}\left [\Delta^{(3)}_{1,1}O_1^{(m)}(i_1,i_2,i_3)+\Delta^{(3)}_{1,2}O_2^{(m)}(i_1,i_2,i_3)+\Delta^{(3)}_{1,3}O_3^{(m)}(i_1,i_2,i_3)\right],\nn
	c^{(m)}_{i_1,i_2+1,i_3}&={1\over (i_2+1)\Delta(1,2,3)}\left [\Delta^{(3)}_{2,1}O_1^{(m)}(i_1,i_2,i_3)+\Delta^{(3)}_{2,2}O_2^{(m)}(i_1,i_2,i_3)+\Delta^{(3)}_{2,3}O_3^{(m)}(i_1,i_2,i_3)\right],\nn
	c^{(m)}_{i_1,i_2,i_3+1}&={1\over (i_3+1)\Delta(1,2,3)}\left [\Delta^{(3)}_{3,1}O_1^{(m)}(i_1,i_2,i_3)+\Delta^{(3)}_{3,2}O_2^{(m)}(i_1,i_2,i_3)+\Delta^{(3)}_{3,3}O_3^{(m)}(i_1,i_2,i_3)\right].     \label{Box-Next-Term}
\end{align}

For a certain rank $m$, by the \textit{inductive assumption} given in the subsection \ref{sec:differential_equation}, the expansion coefficients $c^{(m-1)}_{i_1,i_2,i_3}$ and $c^{(m-1)}_{i_1,i_2}$ are already known in equation \eref{Box-Next-Term},
hence the equation \eref{Box-Next-Term} relates expansion coefficient $c^{(m)}_{i_1+1,i_2,i_3}$, $c^{(m)}_{i_1,i_2+1,i_3}$,$c^{(m)}_{i_1,i_2,i_3+1}$ to $c^{(m)}_{i_1-1,i_2,i_3}$, $c^{(m)}_{i_1,i_2-1,i_3}$,$c^{(m)}_{i_1,i_2,i_3-1}$. 
%In other words, the calculation of expansion coefficients $c^{(m)}_{i_1+1,i_2,i_3}$, $c^{(m)}_{i_1,i_2+1,i_3}$,$c^{(m)}_{i_,i_2,i_3+1}$ is reduced into the calculation of $c^{(m)}_{i_1-1,i_2,i_3}$, $c^{(m)}_{i_1,i_2-1,i_3}$,$c^{(m)}_{i_1,i_2,i_3-1}$ with smaller indices. 
As in previous subsections, according to rank $m$ is even or odd, the expansion coefficient $c^{(m)}_{i_1,i_2,i_3}$ will be reduced into $c^{(m)}_{0,0,0}$ or $c^{(m)}_{1,0,0}$, $c^{(m)}_{0,1,0}$ and $c^{(m)}_{0,0,1}$ finally. As before the other two expansion coefficients $c^{(m)}_{0,1,0}$ and $c^{(m)}_{0,0,1}$ can be got from $c^{(m)}_{1,0,0}$by using the permutation symmetry of  $\{1,2,3\}$.
So the task becomes the determination of the initial expansion coefficient $c^{(m)}_{0,0,0}$ for $m=2r$ or  $c^{(m)}_{1,0,0}$ for $m=2r+1$.

First, we consider the odd rank case with $m=2r+1$. When $i_1=i_2=i_3=0$, equation \eref{Box-Next-Term} becomes
\begin{align}
c^{(2r+1)}_{1,0,0}&=
{1\over \Delta(1,2,3)}\left [\Delta^{(3)}_{1,1}O_1^{(2r+1)}(0,0,0)+\Delta^{(3)}_{1,2}O_2^{(2r+1)}(0,0,0)+\Delta^{(3)}_{1,3}O_3^{(2r+1)}(0,0,0)\right],\nn
c^{(2r+1)}_{0,1,0}&={1\over \Delta(1,2,3)}\left [\Delta^{(3)}_{2,1}O_1^{(2r+1)}(0,0,0)+\Delta^{(3)}_{2,2}O_2^{(2r+1)}(0,0,0)+\Delta^{(3)}_{2,3}O_3^{(2r+1)}(0,0,0)\right],\nn
c^{(2r+1)}_{0,0,1}&={1\over \Delta(1,2,3)}\left [\Delta^{(3)}_{3,1}O_1^{(2r+1)}(0,0,0)+\Delta^{(3)}_{3,2}O_2^{(2r+1)}(0,0,0)+\Delta^{(3)}_{3,3}O_3^{(2r+1)}(0,0,0)\right], \label{Box-Odd-FirstTerm}
\end{align}
which means that we can obtain $c_{1,0,0}^{(2r+1)}$, $c_{0,1,0}^{(2r+1)}$ and $c_{0,0,1}^{(2r+1)}$ from $c_{0,0,0}^{(2r)}$ and $c^{(2r)}_{0,0}$. So we only need to calculate $c_{0,0,0}^{(2r)}$ and from it we can get all expansion coefficients by \eref{Box-Odd-FirstTerm} and \eref{Box-Next-Term} recursively.

Now we consider the computation of $c_{0,0,0}^{(2r)}$. Setting $n=3,i_1=i_2=i_3=0$, the $\cal{T}$-type recurrence relation \eref{recurrence_relation_2}  becomes
\bea
&&2r (D+2r-2)c^{(2r)}_{0,0,0}
+2\b_{12} c^{(2r)}_{1,1,0}+2  \b_{13}c^{(2r)}_{1,0,1}
+2\b_{23}c^{(2r)}_{0,1,1}
+2  \b_{11}c^{(2r)}_{2,0,0}\nn
&&+2
\b_{22}c^{(2r)}_{0,2,0}+2 \b_{33}c^{(2r)}_{0,0,2}\
=8r(2r-1)c^{(2r-2)}_{0,0,0}\label{Box-T}.
\eea
Since the above equation contains six unknown coefficients $c^{(2r)}_{1,1,0},c^{(2r)}_{1,0,1},\cdots,c^{(2r)}_{0,0,2}$ with larger indices, we need to reduce them to expansion coefficients with smaller indices. Then we use \eref{Box-Next-Term} to write expansion coefficients $c^{(2r)}_{1,1,0},c^{(2r)}_{1,0,1},\cdots,c^{(2r)}_{0,0,2}$ as $2r-2$ rank expansion coefficients. After combining the results and \eref{Box-T}, we finally get a recurrence relation of $c_{0,0,0}^{(2r)}$ with respect to $r$ as
\begin{align}
c^{(2r)}_{0,0,0}=&\ {2r-1\over  (D+2 r-5)\Delta(1,2,3)}\Bigg\{\Big[\a_1 \Delta^{(3)}_{1,1}+\a_2 \Delta^{(3)}_{1,2}+\a_3 \Delta^{(3)}_{1,3}\Big]c^{(2r-2)}_{0,0}[\hat{1}]\nn
&\ +\Big[\a_1 \Delta^{(3)}_{2,1}+\a_2 \Delta^{(3)}_{2,2}+\a_3 \Delta^{(3)}_{2,3}\Big]c^{(2r-2)}_{0,0}[\hat{2}]
+\Big[\a_1 \Delta^{(3)}_{3,1}+\a_2 \Delta^{(3)}_{3,2}+\a_3 \Delta^{(3)}_{3,3}\Big]c^{(2r-2)}_{0,0}[\hat{3}]\nn
&\ +\Big[
4 \Delta(1,2,3)-
\a_1^2 \Delta^{(3)}_{1,1}-\a_2^2\Delta^{(3)}_{2,2}-\a_3^2\Delta^{(3)}_{3,3}
-2\a_1\a_2\Delta^{(3)}_{1,2}
-2\a_1\a_3\Delta^{(3)}_{1,3}-2\a_2\a_3\Delta^{(3)}_{2,3}\
\Big]c^{(2r-2)}_{0,0,0}
\Bigg\}\nn
=&\ {2r-1\over  D+2 r-5}\left[(4-\boldsymbol{\a}^TG^{-1}\boldsymbol{\a})c_{0,0,0}^{(2r-2)}+\boldsymbol{\a}^TG^{-1}
\boldsymbol{c}_{0,0}^{(2r-2)}\right],~~~~\label{Box-Even-000Term}
\end{align}
where we have used the same notations as in \eref{sec:tensor triangle}. Using the boundary condition $c^{(0)}_{0,0,0}=0$, it's easy to get the expression for it as
\begin{align}
c^{(2r)}_{0,0,0}= \sum_{i=2}^r \left(\prod_{j=i}^r \frac{2j-1}{2j+D-5}\right) \left(4- \boldsymbol{\alpha}^T  G^{-1} \boldsymbol{\alpha} \right)^{r-i}  \left(\boldsymbol{\alpha}^T  G^{-1} \boldsymbol{c}_{0,0}^{(2i-2)} \right),    \label{Box_expression_even}
\end{align}
where we have considered $(\boldsymbol{c}_0^{(0,0)})^T=(0,0,0)$, so the summation is from $i=2$ to $r$. 

 Since we have obtained  the analytic expression of $c_{0,0,0}^{(2r)}$,   we will show how to obtain all expansion coefficients $c_{i,j,k}^{(m)}$ by using the recurrence relation \eref{Box-Next-Term}, \eref{Box-Odd-FirstTerm} and \eref{Box-Even-000Term} step by step. First we note \eref{Box-Odd-FirstTerm}  can be rewritten as
 \begin{align}
 \left(\begin{array}{l}
 c^{(2r+1)}_{1,0,0} \\
 c^{(2r+1)}_{0,1,0}\\
 c^{(2r+1)}_{0,0,1} \\
 \end{array} \right) = (2r+1)   G^{-1} \left( c_{0,0,0}^{(2r)}~  \boldsymbol{\alpha}  -  \boldsymbol{c}_{0,0}^{(2r)} \right),   \label{Box_expression_odd}
 \end{align}
 Since the expression of $c_{0,0,0}^{(2r)}$ and $\boldsymbol{c}_{0,0}^{(2r)}$ are known, then the results of $c_{1,0,0}^{(2r+1)}$, $c_{0,1,0}^{(2r+1)}$,$c_{0,0,1}^{(2r+1)}$ are easy to get. Second, we will take several examples to illustrate the procedure of calculation. Here for simplicity we denote $\Delta(i_1,i_2,\cdots,i_n;j_1,j_2,\cdots,j_n)$ as the determinant of a $n\times n$  matrix $A$ with entry $A_{ab}=\beta_{i_aj_b}$. Specially, we denote $\Delta(i_1,i_2,\cdots,i_n)\equiv \Delta(i_1,i_2,\cdots,i_n;i_1,i_2,\cdots,i_n)$.
% % % % %
\begin{itemize}
	\item $m<3$: The essence of one-loop reduction is to expand the auxiliary vector $R$ with external momenta to cancel the propagators. To reduce a tensor box integral to the scalar tadpole integral with propagator $P_0$, we need to cancel other three propagators, ie., $P_1,P_2,P_3$, which means three $R$'s are required at least.  So all the expansion coefficients vanish for rank $m<3$. This result can also be obtained from our explicit
recurrence relation \eref{Box_expression_even} and \eref{Box-Next-Term-Matrix}.

	\item $m=3$: We need to determine $c^{(3)}_{1,0,0},c^{(3)}_{1,1,1},c^{(3)}_{1,2,0},c^{(3)}_{3,0,0}$, other expansion coefficients  can be got using the permutation symmetry.
	Using \eref{Box-Next-Term}, we have
	\bea
	c^{(3)}_{1,0,0}&=&{1\over \Delta(1,2,3)}\left [\Delta^{(3)}_{1,1}O_1^{(3)}(0,0,0)+\Delta^{(3)}_{1,2}O_2^{(3)}(0,0,0)+\Delta^{(3)}_{1,3}O_3^{(3)}(0,0,0)\right]\nn
	&=&0,\nn
	c^{(3)}_{1,1,1}&=&{1\over \Delta(1,2,3)}\left [\Delta^{(3)}_{1,1}O_1^{(3)}(0,1,1)+\Delta^{(3)}_{1,2}O_2^{(3)}(0,1,1)+\Delta^{(3)}_{1,3}O_3^{(3)}(0,1,1)\right]\nn
	&=&-\frac{6}{ \Delta (1,2,3)},\nn
	c^{(3)}_{1,2,0}&=&{1\over \Delta(1,2,3)}\left [\Delta^{(3)}_{1,1}O_1^{(3)}(0,2,0)+\Delta^{(3)}_{1,2}O_2^{(3)}(0,2,0)+\Delta^{(3)}_{1,3}O_3^{(3)}(0,2,0)\right]\nn
	&=&\frac{3 \beta _{23}}{\beta _{22} \Delta (1,2,3)}+\frac{3 \beta _{12} \Delta^{(3)} _{1,3}}{\beta _{22} \Delta(1,2) \Delta (1,2,3)},\nn
	c^{(3)}_{3,0,0}&=&{1\over 3\Delta(1,2,3)}\left [\Delta^{(3)}_{1,1}O_1^{(3)}(2,0,0)+\Delta^{(3)}_{1,2}O_2^{(3)}(2,0,0)+\Delta^{(3)}_{1,3}O_3^{(3)}(2,0,0)\right]
	\nn
	&=&\frac{-\beta _{13}\Delta(2,3;1,3) }{\beta _{11} \Delta(1,3)\Delta (1,2,3)}+( 2\leftrightarrow 3),
	\eea
	where we have used
	\bean
	O_1^{(3)}(0,0,0)&=&3\left(\a_1c^{(2)}_{0,0,0}-c^{(2)}_{0,0}[\hat{1}]\right)=0,\nn
	O_2^{(3)}(0,0,0)&=&3\left(\a_2c^{(2)}_{0,0,0}-c^{(2)}_{0,0}[\hat{2}]\right)=0,\nn
	O_3^{(3)}(0,0,0)&=&3\left(\a_3c^{(2)}_{0,0,0}-c^{(2)}_{0,0}[\hat{3}]\right)=0, \eean
\bean	 O_1^{
(3)}(0,1,1)&=&3\left(\a_1c^{(2)}_{0,1,1}-c^{(2)}_{1,1}[\hat{1}]\right)=-\frac{6}{\Delta (2,3)},\nn
	O_2^{(3)}(0,1,1)&=&3\a_2c^{(2)}_{0,1,1}-4c^{(3)}_{0,0,1}=0,\nn
	O_3^{(3)}(0,1,1)&=&3\a_3c^{(2)}_{0,1,1}-4c^{(3)}_{0,1,0}=0,\eean
\bean 
	O_1^{(3)}(0,2,0)&=&3\left(\a_1c^{(2)}_{0,2,0}-c^{(2)}_{2,0}[\hat{1}]\right)=\frac{3 \beta _{23}}{\beta _{22} \Delta(2,3)},\nn
	O_2^{(3)}(0,2,0)&=&3\a_2c^{(2)}_{0,2,0}-2c^{(3)}_{0,1,0}=0,\nn
	O_3^{(3)}(0,2,0)&=&3\left(\a_3c^{(2)}_{0,2,0}-c^{(2)}_{0,2}[\hat{3}]\right)=\frac{3 \beta _{12}}{\beta _{22} \Delta(1,2)},\eean
\bean
	O_1^{(3)}(2,0,0)&=&3\a_1c^{(2)}_{2,0,0}-2c^{(3)}_{1,0,0}=0,\nn
	O_2^{(3)}(2,0,0)&=&3\a_2c^{(2)}_{2,0,0}-3c^{(2)}_{2,0}[\hat{2}]=\frac{3 \beta _{13}}{\beta _{11} \Delta(1,3)},\nn
	O_3^{(3)}(2,0,0)&=&3\a_3c^{(2)}_{2,0,0}-3c^{(2)}_{2,0}[\hat{3}]=\frac{3 \beta _{12}}{\beta _{11} \Delta(1,2)}.
	\eean
	\item $m=4$: We need to determine $c^{(4)}_{0,0,0},c^{(4)}_{1,1,0},c^{(4)}_{2,0,0},c^{(4)}_{1,3,0},c^{(4)}_{2,2,0},c^{(4)}_{4,0,0}$, other expansion coefficients can be got using the permutation symmetry. Using \eref{Box-Even-000Term} we find
	\bea
	c^{(4)}_{0,0,0}=0.
	\eea
	Next using \eref{Box-Next-Term}, we have
	\bea
	c^{(4)}_{1,1,0}&=&{1\over \Delta(1,2,3)}\left [\Delta^{(3)}_{1,1}O_1^{(4)}(0,1,0)+\Delta^{(3)}_{1,2}O_2^{(4)}(0,1,0)+\Delta^{(3)}_{1,3}O_3^{(4)}(0,1,0)\right]\nn
	&=&\frac{12 \alpha_3}{(D-1)  \Delta (1,2,3)}-\left (\frac{12\a_2 \Delta(1,2;1,3)}{(D-1)  \Delta (1,2) \Delta (1,2,3)}+(1\leftrightarrow 2)\right),
	\eea
	\bea
	c^{(4)}_{2,0,0}&=&{1\over 2\Delta(1,2,3)}\left [\Delta^{(3)}_{1,1}O_1^{(4)}(1,0,0)+\Delta^{(3)}_{1,2}O_2^{(4)}(1,0,0)+\Delta^{(3)}_{1,3}O_3^{(4)}(1,0,0)\right] \nn
	&=&\frac{6\left(\alpha _2 \beta _{11}-\alpha _1 \beta _{12}\right)\Delta(2,3;1,2)}{(D-1) \beta _{11} \Delta (1,2) \Delta (1,2,3)}+(2\leftrightarrow 3),
	\eea
	%%%%
	
	%%%%
	\bea
	c^{(4)}_{1,3,0}&=&{1\over \Delta(1,2,3)}\left [\Delta^{(3)}_{1,1}O_1^{(4)}(0,3,0)+\Delta^{(3)}_{1,2}O_2^{(4)}(0,3,0)+\Delta^{(3)}_{1,3}O_3^{(4)}(0,3,0)\right]\nn
	&=&\frac{4 \Delta _{1,3}^{(3)} \Big \{\alpha _2 \beta _{12} \left [(D+1) \beta _{11} \beta _{22}+D \Delta (1,2)\right]+\alpha _1 \beta _{22} \left [(1-D) \beta _{12}^2-2 \beta _{11} \beta _{22}\right]\Big\}}{(D-1) \beta _{22}^2 \Delta (1,2)^2 \Delta (1,2,3)}\nn
	&&+\frac{4  \Big \{\alpha _2 \beta _{23} \left [(D+1) \beta _{22} \beta _{33}+D \Delta (2,3)\right]+\alpha _3 \beta _{22} \left [(1-D) \beta _{23}^2-2 \beta _{22} \beta _{33}\right]\Big\}}{(D-1) \beta _{22}^2 \Delta (2,3) \Delta (1,2,3)}\nn
	&&+{4\Delta_{2,3}^{(3)} \over \beta _{22} \Delta (1,2) \Delta (1,2,3)^2} \left [ \alpha _1 \beta _{12} \Delta _{1,1}^{(3)}+ \alpha _3 \beta _{12} \Delta _{1,3}^{(3)}+\frac{ \Delta _{1,2}^{(3)} \left(2 \alpha _2 \beta _{12}-3 \alpha _1 \beta _{22}+\alpha _2 D \beta _{12}\right)}{(D-1) }\right ]\nn
	&&+{4\Delta_{1,2}^{(3)}\over \beta _{22} \Delta (2,3) \Delta (1,2,3)^2 } \left [\alpha _1 \beta _{23} \Delta _{1,1}^{(3)}+ \alpha _3 \beta _{23} \Delta _{1,3}^{(3)}+\frac{ \Delta _{1,2}^{(3)} \left(2 \alpha _2 \beta _{23}-3 \alpha _3 \beta _{22}+\alpha _2 D \beta _{23}\right)}{(D-1)}\right ],\nn
	\eea
	% % % % % % % % % % %
	\bea
	c^{(4)}_{2,1,1}&=&{1\over 2\Delta(1,2,3)}\left [\Delta^{(3)}_{1,1}O_1^{(4)}(1,1,1)+\Delta^{(3)}_{1,2}O_2^{(4)}(1,1,1)+\Delta^{(3)}_{1,3}O_3^{(4)}(1,1,1)\right]\nn
	&=&\frac{12 \left (\Delta _{1,2}^{(3)}\right)^2 \left(\alpha _3 \beta _{13}-\alpha _1 \beta _{33}\right)}{(D-1) \beta _{33} \Delta (1,3) \Delta (1,2,3)^2}+\frac{12 \left (\Delta _{1,3}^{(3)}\right)^2 \left(\alpha _2 \beta _{12}-\alpha _1 \beta _{22}\right)}{(D-1) \beta _{22} \Delta (1,2) \Delta (1,2,3)^2}
	+\frac{12 \Delta _{1,3}^{(3)} \left(\alpha _2 \beta _{23}-\alpha _3 D \beta _{22}\right)}{(D-1) \beta _{22} \Delta (1,2,3)^2}\nn
	&&+\frac{12 \Delta (2,3) \left(\alpha _3 \beta _{13}-\alpha _1 D \beta _{33}\right)}{(D-1) \beta _{33} \Delta (1,2,3)^2}+\frac{12\Delta _{1,2}^{(3)} \left(2 \alpha _3 \beta _{23}- \alpha _2 (D+1) \beta _{33}\right)}{(D-1) \beta _{33} \Delta (1,2,3)^2},
	\eea
	% % % % % % % % % % %
	\bea
	c^{(4)}_{2,2,0}&=&{1\over 2\Delta(1,2,3)}\left [\Delta^{(3)}_{1,1}O_1^{(4)}(1,2,0)+\Delta^{(3)}_{1,2}O_2^{(4)}(1,2,0)+\Delta^{(3)}_{1,3}O_3^{(4)}(1,2,0)\right]\nn
	&=&{6\Delta _{1,2}^{(3)}\over \Delta(1,2,3)^2}\Bigg \{\frac{ \left(\alpha _2 (D+1) \beta _{23}-2 \alpha _3 \beta _{22}\right)}{(D-1) \beta _{22} }+\frac{ \Delta _{1,3}^{(3)} \left(\alpha _2 (D+1) \beta _{12}-2 \alpha _1 \beta _{22}\right)}{(D-1) \beta _{22} \Delta (1,2) }\Bigg \}\nn
	&&+{6\Delta _{1,1}^{(3)}\over \Delta(1,2,3)^2} \Bigg \{\frac{ \alpha _1 \beta _{23}}{\beta _{22} }+\frac{ \alpha _1 \beta _{12} \Delta _{1,3}^{(3)}}{\beta _{22} \Delta (1,2) }+\frac{\Delta _{2,3}^{(3)} \left( \alpha _2 \beta _{12}- \alpha _1 \beta _{22}\right)}{(D-1) \beta _{22} \Delta (1,2) }+\frac{\Delta _{1,2}^{(3)} \left( \alpha _2 \beta _{23}- \alpha _3 \beta _{22}\right)}{(D-1) \beta _{22} \Delta (2,3) }\Bigg\}\nn
	&&+{6\Delta _{1,3}^{(3)}\over \Delta(1,2,3)^2}\Bigg \{
	\frac{ \alpha _3 \beta _{12} \Delta _{1,3}^{(3)}}{\beta _{22} \Delta (1,2) }+\frac{ \alpha _3 \beta _{23}}{\beta _{22} }-\frac{ \Delta (1,2,3)\Big [\alpha _2 D \Delta (1,2)+(D+1) \beta _{12} \left(\alpha _1 \beta _{22}-\alpha _2 \beta _{12}\right)\Big ]}{(D-1) \beta _{22} \Delta (1,2)^2 }\Bigg\},\nn
	\eea
	%%%%
	
	%%%%
	\bea
	c^{(4)}_{4,0,0}&=&{1\over 4\Delta(1,2,3)}\left [\Delta^{(3)}_{1,1}O_1^{(4)}(3,0,0)+\Delta^{(3)}_{1,2}O_2^{(4)}(3,0,0)+\Delta^{(3)}_{1,3}O_3^{(4)}(3,0,0)\right]\nn
	&=&\frac{\alpha _2 \beta _{13} \left(\Delta _{1,2}^{(3)}\right)^2}{\beta _{11} \Delta (1,3) \Delta (1,2,3)^2}+\frac{\alpha _3 \beta _{12} \left (\Delta _{1,3}^{(3)}\right)^2}{\beta _{11} \Delta (1,2) \Delta (1,2,3)^2}
	+ \frac{\Delta _{1,2}^{(3)}\Delta _{1,3}^{(3)} \Big [\alpha _3 \beta _{13} \Delta (1,2)+\alpha _2 \beta _{12} \Delta (1,3)\Big]}{\beta _{11} \Delta (1,2) \Delta (1,3) \Delta (1,2,3)^2}\nn
	&&+\Delta _{1,2}^{(3)}\left \{\frac{\alpha _1 \beta _{13} \left [(D+1) \beta _{11} \beta _{33}+D \Delta (1,3)\right]+\alpha _3 \beta _{11} \left [-2 \beta _{11} \beta _{33}-(D-1) \beta _{13}^2\right]}{(D-1) \beta _{11}^2 \Delta (1,3)^2 \Delta (1,2,3)}\right\}\nn
	&&+\Delta_{1,1}^{(3)} \left \{\frac{\Delta _{1,2}^{(3)} \left(\alpha _1 (D+2) \beta _{13}-3 \alpha _3 \beta _{11}\right)}{(D-1) \beta _{11} \Delta (1,3) \Delta (1,2,3)^2}+\frac{\Delta _{1,3}^{(3)} \left(\alpha _1 (D+2) \beta _{12}-3 \alpha _2 \beta _{11}\right)}{(D-1) \beta _{11} \Delta (1,2) \Delta (1,2,3)^2}\right\}\nn
	&&+\frac{\Delta _{1,3}^{(3)} \Big  \{\alpha _1 \beta _{12} \left [(D+1) \beta _{11} \beta _{22}+D \Delta (1,2)\right]+\alpha _2 \beta _{11} \left [-2 \beta _{11} \beta _{22}-(D-1) \beta _{12}^2\right]\Big\}}{(D-1) \beta _{11}^2 \Delta (1,2)^2 \Delta (1,2,3)},
	\eea
	where  we have used
	\bea
	O_1^{(4)}(0,1,0)&=&4\left(\a_1c^{(3)}_{0,1,0}-c^{(3)}_{1,0}[\hat{1}]\right)=\frac{12 \left(\alpha _3 \beta _{22}-\alpha _2 \beta _{23}\right)}{(D-1) \beta _{22} \Delta (2,3)},\nn
	O_2^{(4)}(0,1,0)&=&4\a_2c^{(3)}_{0,1,0}-4c^{(4)}_{0,0,0}=0,\nn
	O_3^{(4)}(0,1,0)&=&4\left(\a_3c^{(3)}_{0,1,0}-c^{(3)}_{0,1}[\hat{3}]\right)=\frac{12 \left(\alpha _1 \beta _{22}-\alpha _2 \beta _{12}\right)}{(D-1) \beta _{22} \Delta (1,2)},\nonumber
	\eea
	\bea
	O_1^{(4)}(1,0,0)&=&4\a_1c^{(3)}_{1,0,0}-4c^{(4)}_{0,0,0}=0,\nn
	O_2^{(4)}(1,0,0)&=&4\left(\a_2c^{(3)}_{1,0,0}-c^{(3)}_{1,0}[\hat{2}]\right)=\frac{12 \left(\alpha _3 \beta _{11}-\alpha _1 \beta _{13}\right)}{(D-1) \beta _{11} \Delta (1,3)},\nn
	O_3^{(4)}(1,0,0)&=&4\left(\a_3c^{(3)}_{1,0,0}-c^{(3)}_{1,0}[\hat{3}]\right)=\frac{12 \left(\alpha _2 \beta _{11}-\alpha _1 \beta _{12}\right)}{(D-1) \beta _{11} \Delta (1,2)},\nonumber
	\eea
	\begin{align}
	O_1^{(4)}(0,3,0)=&\ 4\left(\a_1c^{(3)}_{0,3,0}-c^{(3)}_{3,0}[\hat{1}]\right)\nn
	=&\ \frac{-4\alpha _3 \left [2 \beta _{22} \beta _{33}+ (D-1) \beta _{23}^2\right]}{(D-1) \beta _{22} \Delta (2,3)^2}+\frac{4 \alpha _2 \beta _{23} \left [(D+1) \beta _{22} \beta _{33}
		+D \Delta (2,3)\right]}{(D-1) \beta _{22}^2 \Delta (2,3)^2}\nn
	&+\frac{4 \alpha _1 \beta _{12} \Delta_{2,3}^{(3)}}{\beta _{22} \Delta (1,2) \Delta (1,2,3)}+\frac{4 \alpha _1 \beta _{23} \Delta_{1,2}^{(3)}}{\beta _{22} \Delta (2,3) \Delta (1,2,3)},\no
	\end{align}
	\begin{align}
	O_2^{(4)}(0,3,0)&=4\a_2c^{(3)}_{0,3,0}-2c^{(4)}_{0,2,0}\nn
	&=\frac{4 \Delta_{2,3}^{(3)} \left(2 \alpha _2 \beta _{12}-3 \alpha _1 \beta _{22}+\alpha _2 D \beta _{12}\right)}{(D-1) \beta _{22} \Delta (1,2) \Delta (1,2,3)}+\frac{4 \Delta_{1,2}^{(3)}  \left(-3 \alpha _3 \beta _{22}+2 \alpha _2 \beta _{23}+\alpha _2 D \beta _{23}\right)}{(D-1) \beta _{22} \Delta (2,3) \Delta (1,2,3)},\no
	\end{align}
	\begin{align}
	O_3^{(4)}(0,3,0)&=4\left(\a_3c^{(3)}_{0,3,0}-c^{(3)}_{0,3}[\hat{3}]\right)\nn
	&=\frac{-4\alpha _1 \left [2\beta _{11} \beta _{22}+ (D-1) \beta _{12}^2\right]}{(D-1) \beta _{22} \Delta (1,2)^2}+\frac{4 \alpha _2 \beta _{12} \left [(D+1) \beta _{11} \beta _{22}+D \Delta (1,2)\right]}{(D-1) \beta _{22}^2 \Delta (1,2)^2}\nn
	&\ \ \ +\frac{4 \alpha _3 \beta _{12} \Delta_{2,3}^{(3)}}{\beta _{22} \Delta (1,2) \Delta (1,2,3)}+\frac{4 \alpha _3 \beta _{23} \Delta_{1,2}^{(3)}}{\beta _{22} \Delta (2,3) \Delta (1,2,3)},\no
	\end{align}
	\bea
	O_1^{(4)}(1,1,1)&=&4\a_1c^{(3)}_{1,1,1}-4c^{(4)}_{0,1,1}\nn
	&=&\frac{24 \left(\alpha _3 \beta _{13}-\alpha _1 D \beta _{33}\right)}{(D-1) \beta _{33} \Delta (1,2,3)}+\frac{24 \Delta _{1,2}^{(3)} \left(\alpha _3 \beta _{23}-\alpha _2 \beta _{33}\right)}{(D-1) \beta _{33} \Delta (2,3) \Delta (1,2,3)},\nn
	O_2^{(4)}(1,1,1)&=&4\a_2c^{(3)}_{1,1,1}-4c^{(4)}_{1,0,1}\nn
	&=&\frac{24 \left(\alpha _3 \beta _{23}-\alpha _2 D \beta _{33}\right)}{(D-1) \beta _{33} \Delta (1,2,3)}+\frac{24 \Delta _{1,2}^{(3)} \left(\alpha _3 \beta _{13}-\alpha _1 \beta _{33}\right)}{(D-1) \beta _{33} \Delta (1,3) \Delta (1,2,3)},\nn
	O_3^{(4)}(1,1,1)&=&4\a_3c^{(3)}_{1,1,1}-4c^{(4)}_{1,1,0}\nn
	&=&\frac{24 \left(\alpha _2 \beta _{23}-\alpha _3 D \beta _{22}\right)}{(D-1) \beta _{22} \Delta (1,2,3)}+\frac{24 \Delta _{1,3}^{(3)} \left(\alpha _2 \beta _{12}-\alpha _1 \beta _{22}\right)}{(D-1) \beta _{22} \Delta (1,2) \Delta (1,2,3)},\no
	\eea
	\bea
	O_1^{(4)}(1,2,0)&=&4\a_1c^{(3)}_{1,2,0}-2c^{(4)}_{0,2,0}\nn
	&=&\frac{12 \alpha _1 \beta _{23}}{\beta _{22} \Delta (1,2,3)}+\frac{12 \alpha _1 \beta _{12} \Delta _{1,3}^{(3)}}{\beta _{22} \Delta (1,2) \Delta (1,2,3)}+\frac{12 \Delta _{2,3}^{(3)} \left(\alpha _2 \beta _{12}-\alpha _1 \beta _{22}\right)}{(D-1) \beta _{22} \Delta (1,2) \Delta (1,2,3)}\nn
	&&+\frac{12 \Delta _{1,2}^{(3)} \left(\alpha _2 \beta _{23}-\alpha _3 \beta _{22}\right)}{(D-1) \beta _{22} \Delta (2,3) \Delta (1,2,3)},\no
	\eea
	\bea
	O_2^{(4)}(1,2,0)&=&4\a_2c^{(3)}_{1,2,0}-2c^{(4)}_{1,1,0}\nn
	&=&\frac{12 \left(\alpha _2 (D+1) \beta _{23}-2 \alpha _3 \beta _{22}\right)}{(D-1) \beta _{22} \Delta (1,2,3)}+\frac{12 \Delta _{1,3}^{(3)} \left(\alpha _2 (D+1) \beta _{12}-2 \alpha _1 \beta _{22}\right)}{(D-1) \beta _{22} \Delta (1,2) \Delta (1,2,3)},\no
	\eea
	\bea
	O_3^{(4)}(1,2,0)&=&4\left(\a_3c^{(3)}_{1,2,0}-c^{(3)}_{1,2}[\hat{3}]\right)\nn
	&=&\frac{12 \alpha _3 \beta _{23} \Delta (1,2)+12\beta _{12} \Delta_{1,3}^{(3)}}{\beta _{22} \Delta (1,2) \Delta (1,2,3)}
	+\frac{12 (D+1) \beta _{12} \left(\alpha _1 \beta _{22}-\alpha _2 \beta _{12}\right)-12\alpha_2D\Delta(1,2)}{(D-1) \beta _{22} \Delta (1,2)^2},\no
	\eea
	\bea
	O_1^{(4)}(3,0,0)&=&4\a_1c^{(3)}_{3,0,0}-2c^{(4)}_{2,0,0}\nn
	&=&\frac{4 \Delta _{1,2}^{(3)} \left(\alpha _1 (D+2) \beta _{13}-3 \alpha _3 \beta _{11}\right)}{(D-1) \beta _{11} \Delta (1,3) \Delta (1,2,3)}+\frac{4 \Delta _{1,3}^{(3)} \left(\alpha _1 (D+2) \beta _{12}-3 \alpha _2 \beta _{11}\right)}{(D-1) \beta _{11} \Delta (1,2) \Delta (1,2,3)},\no
	\eea
	\bea
	O_2^{(4)}(3,0,0)&=&4\left(\a_2c^{(3)}_{3,0,0}-c^{(3)}_{3,0}[\hat{2}]\right)\nn
	&=&\frac{4 \alpha _2 \beta _{13} \Delta _{1,2}^{(3)}}{\beta _{11} \Delta (1,3) \Delta (1,2,3)}+\frac{4 \alpha _2 \beta _{12} \Delta _{1,3}^{(3)}}{\beta _{11} \Delta (1,2) \Delta (1,2,3)}\nn
	&&+\frac{4 \Big[\alpha _1 \beta _{13} \left((D+1) \beta _{11} \beta _{33}+D \Delta (1,3)\right)+\alpha _3 \beta _{11} \left(-2 \beta _{11} \beta _{33}-(D-1) \beta _{13}^2\right)\Big]}{(D-1) \beta _{11}^2 \Delta (1,3)^2},\no
	\eea
	\bea
	O_3^{(4)}(3,0,0)&=&4\left(\a_3c^{(3)}_{3,0,0}-c^{(3)}_{3,0}[\hat{3}]\right)\nn
	&=&\frac{4 \alpha _3 \beta _{13} \Delta _{1,2}^{(3)}}{\beta _{11} \Delta (1,3) \Delta (1,2,3)}+\frac{4 \alpha _3 \beta _{12} \Delta _{1,3}^{(3)}}{\beta _{11} \Delta (1,2) \Delta (1,2,3)}\nn
	&&+\frac{4 \Big[\alpha _1 \beta _{12} \left((D+1) \beta _{11} \beta _{22}+D \Delta (1,2)\right)+\alpha _2 \beta _{11} \left(-2 \beta _{11} \beta _{22}-(D-1) \beta _{12}^2\right)\Big]}{(D-1) \beta _{11}^2 \Delta (1,2)^2}.\no
	\eea
\end{itemize} 	
% % % % %
The above examples are enough to illustrate the procedure of calculating expansion coefficients $c_{i,j,k}^{(m)}$.
Rough speaking, we calculate the expansion coefficients $c_{i,j,k}^{(m)}$ from lower rank to higher rank, and at a fixed rank $m$, we prefer to calculate the expansion coefficients $c_{i,j,k}^{(m)}$ with smaller  $i+j+k$ first. For a general rank $m_0$, we need to calculate all expansion coefficients with rank $m<m_0$ to calculate the tadpole coefficient with rank $m_0$.
The algorithm is the same as previous ones.\footnote{It is similar to the tensor triangle case, but due to there are three indices in $c_{i,j,k}^{(m)}$, we don't draw a picture to illustrate the algorithm.}
\begin{itemize}
	
	\item Step 1: For rank $m<3$, we have shown that all expansion coefficients vanish.
	\item Step 2: Consider the rank $m=3$, calculate the expansion coefficients $c^{(3)}_{1,0,0}, c^{(3)}_{0,1,0}$, $c^{(3)}_{0,0,1}$,$\cdots,c^{(3)}_{3,0,0}$ by \eref{Box-Next-Term} as we illustrated before.
	\item Step 3: Consider the rank $m=4$,  combining the permutation symmetry, calculate the expansion coefficients $c^{(4)}_{0,0,0}, c^{(4)}_{1,1,0},c^{(4)}_{2,0,0}$, $\dots, c^{(4)}_{4,0,0}$  successively by \eref{Box-Even-000Term} and \eref{Box-Next-Term} as illustrated before.
	
	$\cdots$
	
	\item Step $m \le m_0$: Combine the permutation symmetry,
	if $m=2r$, calculate $c_{0,0,0}^{(2r)}$,$ c_{1,1,0}^{(2r)}$,\\$c_{2,0,0}^{(2r)}$,$\cdots$,$c_{2r,0,0}^{(2r)}$ successively \eref{Box-Even-000Term} and \eref{Box-Next-Term}.
	if $m=2r+1$, calculate $c_{1,0,0}^{(2r+1)}, c_{1,1,1}^{(2r+1)}$,\\$c_{1,2,0}^{(2r+1)},c_{3,0,0}^{(2r+1)}, \cdots,c_{2r+1,0,0}^{(2r+1)}$ successively by using \eref{Box-Next-Term}.
	\item Final step: Combine all expansion coefficients to get the tadpole coefficient by \eref{Box-Exp}.
\end{itemize}
With the help of Mathematica, it is easy to implement recurrence relation \eref{Box_expression_even} and \eref{Box-Next-Term-Matrix} to  automatically produce analytic expression for reduction coefficients of any rank. 
\subsection{Tadpole coefficients of tensor pentagon integral}

% % % % % % % % % % % % % %
At last, we consider the fourth nontrivial case, $n=4$ for \eref{Idea-1-1}, i.e., tensor pentagon integral. In this case, the general form of tadpole coefficients \eref{C0-exp-1} can be written as
\bea
C_1^{(0)}(m)=\sum_{\{i_1,i_2,i_3,i_4\}}(M_0^2)^{-4}(M_0^2s_{00})^{m-i_1-i_2-i_3-i_4\over 2}c^{(m)}_{i_1,i_2,i_3,i_4}s_{01}^{i_1}s_{02}^{i_2}s_{03}^{i_3}s_{04}^{i_4}.\label{Pen-Exp}
\eea
As a result, the $\cal{D}$-type recurrence relations are given by setting  $n=4$ in  \eref{recurrence_relation_1}:
\bea
&&(m+1-i_1-i_2-i_3-i_4)c^{(m)}_{i_1-1,i_2,i_3,i_4}
+(i_1+1)\b_{11}c^{(m)}_{i_1+1,i_2,i_3,i_4}+(i_2+1)\b_{12}c^{(m)}_{i_1,i_2+1,i_3,i_4}\nn
&&+(i_3+1)\b_{13}c^{(m)}_{i_1,i_2,i_3+1,i_4}
+(i_4+1)\b_{14}c^{(m)}_{i_1,i_2,i_3,i_4+1}=m\a_1c^{(m-1)}_{i_1,i_2,i_3,i_4}-m\delta_{i_1,0}c^{(m-1)}_{i_2,i_3,i_4}[\hat{1}],\nn
&&(m+1-i_1-i_2-i_3-i_4)c^{(m)}_{i_1,i_2-1,i_3,i_4}
+(i_1+1)\b_{12}c^{(m)}_{i_1+1,i_2,i_3,i_4}+(i_2+1)\b_{22}c^{(m)}_{i_1,i_2+1,i_3,i_4}\nn
&&+(i_3+1)\b_{23}c^{(m)}_{i_1,i_2,i_3+1,i_4}
+(i_4+1)\b_{24}c^{(m)}_{i_1,i_2,i_3,i_4+1}=m\a_2c^{(m-1)}_{i_1,i_2,i_3,i_4}-m\delta_{i_2,0}c^{(m-1)}_{i_1,i_3,i_4}[\hat{2}],\nn
&&(m+1-i_1-i_2-i_3-i_4)c^{(m)}_{i_1,i_2,i_3-1,i_4}
+(i_1+1)\b_{13}c^{(m)}_{i_1+1,i_2,i_3,i_4}+(i_2+1)\b_{23}c^{(m)}_{i_1,i_2+1,i_3,i_4}\nn
&&+(i_3+1)\b_{33}c^{(m)}_{i_1,i_2,i_3+1,i_4}
+(i_4+1)\b_{34}c^{(m)}_{i_1,i_2,i_3,i_4+1}=m\a_3c^{(m-1)}_{i_1,i_2,i_3,i_4}-m\delta_{i_3,0}c^{(m-1)}_{i_1,i_2,i_4}[\hat{3}],\nn
&&(m+1-i_1-i_2-i_3-i_4)c^{(m)}_{i_1,i_2,i_3,i_4-1}
+(i_1+1)\b_{14}c^{(m)}_{i_1+1,i_2,i_3,i_4}+(i_2+1)\b_{24}c^{(m)}_{i_1,i_2+1,i_3,i_4}\nn
&&+(i_3+1)\b_{34}c^{(m)}_{i_1,i_2,i_3+1,i_4}
+(i_4+1)\b_{44}c^{(m)}_{i_1,i_2,i_3,i_4+1}=m\a_4c^{(m-1)}_{i_1,i_2,i_3,i_4}-m\delta_{i_4,0}c^{(m-1)}_{i_1,i_2,i_3}[\hat{4}],
\eea
These equations can be rewritten in a more compact form as 
where $[\hat{i}]$ means the $i$-th propagator is removed from the  propagators $P_0,P_1,P_2,P_3,P_4$. These equations can be rewritten in a more compact form as 
\begin{align}
G\left(\begin{array}{l}
(i_1+1)c^{(m)}_{i_1+1,i_2,i_3,i_4} \\
(i_2+1)c^{(m)}_{i_1,i_2+1,i_3,i_4}\\
(i_3+1)c^{(m)}_{i_1,i_2,i_3+1,i_4}\\
(i_4+1)c^{(m)}_{i_1,i_2,i_3,i_4+1}\\
\end{array} \right) =     \boldsymbol{O}^{(m)}(i_1,i_2,i_3,i_4) ,   \label{Pen-Di-Matrix}
\end{align}
where we have defined the vector  $\boldsymbol{O}^{(m)}(i_1,i_2,i_3,i_4)^T=(O_1^{(m)}(i_1,i_2,i_3,i_4),O_2^{(m)}(i_1,i_2,i_3,i_4)$, $O_3^{(m)}(i_1,i_2,i_3,i_4)$, $O_4^{(m)}(i_1,i_2,i_3,i_4))$ as

\bea
O_1^{(m)}(i_1,i_2,i_3,i_4)&=&m\left(\a_1c^{(m-1)}_{i_1,i_2,i_3,i_4}-\delta_{0,i_1}c^{(m-1)}_{i_2,i_3,i_4}[\hat{1}]\right)-(m+1-i_1-i_2-i_3-i_4)c^{(m)}_{i_1-1,i_2,i_3,i_4},\nn
O_2^{(m)}(i_1,i_2,i_3,i_4)&=&m\left(\a_2c^{(m-1)}_{i_1,i_2,i_3,i_4}-\delta_{0,i_2}c^{(m-1)}_{i_1,i_3,i_4}[\hat{2}]\right)-(m+1-i_1-i_2-i_3-i_4)c^{(m)}_{i_1,i_2-1,i_3,i_4},\nn
O_3^{(m)}(i_1,i_2,i_3,i_4)&=&m\left(\a_3c^{(m-1)}_{i_1,i_2,i_3,i_4}-\delta_{0,i_3}c^{(m-1)}_{i_1,i_2,i_4}[\hat{3}]\right)-(m+1-i_1-i_2-i_3-i_4)c^{(m)}_{i_1,i_2,i_3-1,i_4},\nn
O_4^{(m)}(i_1,i_2,i_3,i_4)&=&m\left(\a_4c^{(m-1)}_{i_1,i_2,i_3,i_4}-\delta_{0,i_4}c^{(m-1)}_{i_1,i_2,i_3}[\hat{4}]\right)-(m+1-i_1-i_2-i_3-i_4)c^{(m)}_{i_1,i_2,i_3,i_4-1}.\nonumber
\eea
Then we can solve $c^{(m)}_{i_1+1,i_2,i_3,i_4},c^{(m)}_{i_1,i_2+1,i_3,i_4},c^{(m)}_{i_1,i_2,i_3+1,i_4},c^{(m)}_{i_1,i_2,i_3,i_4+1}$ as
\begin{align}
\left(\begin{array}{l}
c^{(m)}_{i_1+1,i_2,i_3,i_4} \\
c^{(m)}_{i_1,i_2+1,i_3,i_4}\\
c^{(m)}_{i_1,i_2,i_3+1,i_4}\\
c^{(m)}_{i_1,i_2,i_3,i_4+1}\\
\end{array} \right) = \left(\begin{array}{llll}
{1\over i_1+1}&0&0&0\\
0&{1\over i_2+1}&0&0\\
0&0&{1\over i_3+1}&0\\
0&0&0&{1\over i_4+1}
\end{array} \right)   G^{-1}   \boldsymbol{O}^{(m)}(i_1,i_2,i_3,i_4) .   \label{Pen-Next-Term-Matrix}
\end{align}
More explicitly, they are
\bea
c^{(m)}_{i_1+1,i_2,i_3,i_4}&=&{1\over (i_1+1)\Delta(1,2,3,4)}\Big[\Delta^{(4)}_{1,1}O_1^{(m)}(i_1,i_2,i_3,i_4)+\Delta^{(4)}_{1,2}O_2^{(m)}(i_1,i_2,i_3,i_4)\nn
&&+\Delta^{(4)}_{1,3}O_3^{(m)}(i_1,i_2,i_3,i_4)+\Delta^{(4)}_{1,4}O_4^{(m)}(i_1,i_2,i_3,i_4)\Big],\nn
c^{(m)}_{i_1,i_2+1,i_3,i_4}&=&{1\over (i_2+1)\Delta(1,2,3,4)}\Big[\Delta^{(4)}_{2,1}O_1^{(m)}(i_1,i_2,i_3,i_4)+\Delta^{(4)}_{2,2}O_2^{(m)}(i_1,i_2,i_3,i_4)\nn
&&+\Delta^{(4)}_{2,3}O_3^{(m)}(i_1,i_2,i_3,i_4)+\Delta^{(4)}_{2,4}O_4^{(m)}(i_1,i_2,i_3,i_4)\Big],\nn
c^{(m)}_{i_1,i_2,i_3+1,i_4}&=&{1\over (i_3+1)\Delta(1,2,3,4)}\Big[\Delta^{(4)}_{3,1}O_1^{(m)}(i_1,i_2,i_3,i_4)+\Delta^{(4)}_{3,2}O_2^{(m)}(i_1,i_2,i_3,i_4)\nn
&&+\Delta^{(4)}_{3,3}O_3^{(m)}(i_1,i_2,i_3,i_4)+\Delta^{(4)}_{3,4}O_4^{(m)}(i_1,i_2,i_3,i_4)\Big],\nn
c^{(m)}_{i_1,i_2,i_3,i_4+1}&=&{1\over (i_4+1)\Delta(1,2,3,4)}\Big[\Delta^{(4)}_{4,1}O_1^{(m)}(i_1,i_2,i_3,i_4)+\Delta^{(4)}_{4,2}O_2^{(m)}(i_1,i_2,i_3,i_4)\nn
&&+\Delta^{(4)}_{4,3}O_3^{(m)}(i_1,i_2,i_3,i_4)+\Delta^{(4)}_{4,4}O_4^{(m)}(i_1,i_2,i_3,i_4)\Big].
~~~~\label{Pen-Next-Term}\eea

For a certain rank $m$, by the \textit{inductive assumption} given in the subsection \ref{sec:differential_equation}, the expansion coefficients $\{c^{(m-1)}_{i,j,k,l}, c^{(m-1)}_{i,j,k}\},\forall i,j,k,l$, are already known in equation \eref{Pen-Next-Term}.
Then equation \eref{Pen-Next-Term} relates expansion coefficient $c^{(m)}_{i+1,j,k,l}$, $c^{(m)}_{i,j+1,k,l}$,$c^{(m)}_{i,j,k+1,l}$ ,$c^{(m)}_{i,j,k,l+1}$ with $c^{(m)}_{i-1,j,k,l}$, $c^{(m)}_{i,j-1,k,l}$,$c^{(m)}_{i,j,k-1,l}$,$c^{(m)}_{i,j,k,l-1}$, in other words, the calculation of expansion coefficients $c^{(m)}_{i+1,j,k,l}$, $c^{(m)}_{i,j+1,k,l}$,$c^{(m)}_{i,j,k+1,l}$ ,$c^{(m)}_{i,j,k,l+1}$ is reduced into the calculation of $c^{(m)}_{i-1,j,k,l}$, $c^{(m)}_{i,j-1,k,l}$,$c^{(m)}_{i,j,k-1,l}$,$c^{(m)}_{i,j,k,l-1}$.
Then according to $m$ is even or odd, the expansion coefficient $c^{(m)}_{i,j,k,l}$ is reduced into $c^{(2r)}_{0,0,0,0}$ or $c^{(2r+1)}_{1,0,0,0}$.\footnote{Where $c^{(2r+1)}_{0,1,0,0}$,$c^{(2r+1)}_{0,0,1,0}$,$c^{(2r+1)}_{0,0,0,1}$ can be got by using the permutation symmetry. }
Now the task becomes the determination of the initial expansion coefficient $c^{(2r)}_{0,0,0,0}$ for $m=2r$ or $c^{(2r+1)}_{1,0,0,0}$ for $m=2r+1$.

First, we consider the odd rank case with $m=2r+1$. When $m=2r+1,i_1=i_2=i_3=i_4=0$, equation \eref{Pen-Next-Term} becomes
\bea
c^{(2r+1)}_{1,0,0,0}&=&{1\over \Delta(1,2,3,4)}\Big[\Delta^{(4)}_{1,1}O_1^{(2r+1)}(0,0,0,0)+\Delta^{(4)}_{1,2}O_2^{(2r+1)}(0,0,0,0)\nn
&&+\Delta^{(4)}_{1,3}O_3^{(2r+1)}(0,0,0,0)+\Delta^{(4)}_{1,4}O_4^{(2r+1)}(0,0,0,0)\Big],\nn
c^{(2r+1)}_{0,1,0,0}&=&{1\over \Delta(1,2,3,4)}\Big[\Delta^{(4)}_{2,1}O_1^{(2r+1)}(0,0,0,0)+\Delta^{(4)}_{2,2}O_2^{(2r+1)}(0,0,0,0)\nn
&&+\Delta^{(4)}_{2,3}O_3^{(2r+1)}(0,0,0,0)+\Delta^{(4)}_{2,4}O_4^{(2r+1)}(0,0,0,0)\Big],\nn
c^{(2r+1)}_{0,0,1,0}&=&{1\over \Delta(1,2,3,4)}\Big[\Delta^{(4)}_{3,1}O_1^{(2r+1)}(0,0,0,0)+\Delta^{(4)}_{3,2}O_2^{(2r+1)}(0,0,0,0)\nn
&&+\Delta^{(4)}_{3,3}O_3^{(2r+1)}(0,0,0,0)+\Delta^{(4)}_{3,4}O_4^{(2r+1)}(0,0,0,0)\Big],\nn
c^{(2r+1)}_{0,0,0,1}&=&{1\over \Delta(1,2,3,4)}\Big[\Delta^{(4)}_{4,1}O_1^{(2r+1)}(0,0,0,0)+\Delta^{(4)}_{4,2}O_2^{(2r+1)}(0,0,0,0)\nn
&&+\Delta^{(4)}_{4,3}O_3^{(2r+1)}(0,0,0,0)+\Delta^{(4)}_{4,4}O_4^{(2r+1)}(0,0,0,0)\Big]
,~~~~\label{Pen-Odd-FirstTerm}\eea
which means that we can obtain $c_{1,0,0,0}^{(2r+1)}$, $c_{0,1,0,0}^{(2r+1)}$  $c_{0,0,1,0}^{(2r+1)}$,
$c_{0,0,0,1}^{(2r+1)}$ from $c_{0,0,0,0}^{(2r)}$ and $c^{(2r)}_{0,0,0}$. So we only need to calculate $c_{0,0,0,0}^{(2r)}$, then from $c_{0,0,0,0}^{(2r)}$ we can get all expansion coefficients by \eref{Pen-Odd-FirstTerm} and \eref{Pen-Next-Term} recursively.

Now we consider the computation of $c_{0,0,0,0}^{(2r)}$. When $n=4,i_1=i_2=i_3=i_4=0$, the $\cal{T}$-type relation \eref{recurrence_relation_2} becomes
\bea
&&4r(2r-1)c^{(2r-2)}_{0,0,0,0}=r (D+2r-2)c^{(2r)}_{0,0,0,0}
+ \b_{12} c^{(2r)}_{1,1,0,0}+ \b_{13}c^{(2r)}_{1,0,1,0}+  \b_{14}c^{(2r)}_{1,0,0,1}\nn
&+& \b_{23}c^{(2r)}_{0,1,1,0}
+\b_{34}c^{(2r)}_{0,0,1,1}+ \b_{11}c^{(2r)}_{2,0,0,0}
+
\b_{22}c^{(2r)}_{0,2,0,0}+\b_{33}c^{(2r)}_{0,0,2,0}+ \b_{44}c^{(2r)}_{0,0,0,2},~~~\label{Pen-T}
\eea
Since the above equation contains unknown expansion coefficients $c^{(2r)}_{1,1,0,0},c^{(2r)}_{1,0,1,0},\cdots,c^{(2r)}_{0,0,0,2}$,  we use \eref{Pen-Next-Term} to write them as $2r-2$ rank expansion coefficients. Combining these equations and \eref{Pen-T}, we finally get the recurrence relation of  $c^{(2r)}_{0,0,0,0}$  as
\bea
c^{(2r)}_{0,0,0,0}&=&{2r-1\over  (D+2 r-6)\Delta(1,2,3,4)}\Bigg \{\Big [4 \Delta(1,2,3,4)
-\a_1^2 \Delta^{(4)}_{1,1}-\a_2^2\Delta^{(4)}_{2,2}-\a_3^2\Delta^{(4)}_{3,3}
-\a_4^2\Delta^{(4)}_{4,4}\nn
&&-2\a_1\a_2\Delta^{(4)}_{1,2}
-2\a_1\a_3\Delta^{(4)}_{1,3}-2\a_1\a_4\Delta^{(4)}_{1,4}
-2\a_2\a_3\Delta^{(4)}_{2,3}-2\a_3\a_4\Delta^{(4)}_{3,4}\Big]c^{(2r-2)}_{0,0,0,0}\nn
&&+\big(\a_1 \Delta^{(4)}_{1,1}+\a_2 \Delta^{(4)}_{1,2}+\a_3\Delta^{(4)}_{1,3}+\a_4\Delta^{(4)}_{1,4}\big)c^{(2r-2)}_{0,0,0}[\hat{1}]\nn
&&+\big(\a_1 \Delta^{(4)}_{2,1}+\a_2 \Delta^{(4)}_{2,2}+\a_3\Delta^{(4)}_{2,3}+\a_4\Delta^{(4)}_{2,4}\big)c^{(2r-2)}_{0,0,0}[\hat{2}]\nn
&&+\big(\a_1 \Delta^{(4)}_{3,1}+\a_2 \Delta^{(4)}_{3,2}+\a_3\Delta^{(4)}_{3,3}+\a_4\Delta^{(4)}_{3,4}\big)c^{(2r-2)}_{0,0,0}[\hat{3}]\nn
&&+\big(\a_1 \Delta^{(4)}_{4,1}+\a_2 \Delta^{(4)}_{4,2}+\a_3\Delta^{(4)}_{4,3}+\a_4\Delta^{(4)}_{4,4}\big)c^{(2r-2)}_{0,0,0}[\hat{4}]
\Bigg\}\nn
&=&{2r-1\over  D+2 r-6}\left[(4-\boldsymbol{\a}^TG^{-1}\boldsymbol{\a})c_{0,0,0,0}^{(2r-2)}+\boldsymbol{\a}^TG^{-1}\boldsymbol{c}_{0,0,0}^{(2r-2)}\right],~~~~\label{Pen-Even-0000Term}
\eea
where we have used the denotations in \eref{sec:tensor triangle}.
Using the boundary condition $c_{0,0,0,0}^{(0)}=0$, it's easy to get the expression for $c^{(2r)}_{0,0,0,0}$ as
\begin{align}
c^{(2r)}_{0,0,0,0}= \sum_{i=2}^r \left(\prod_{j=i}^r \frac{2j-1}{2j+D-6}\right) \left(4- \boldsymbol{\alpha}^T  G^{-1} \boldsymbol{\alpha} \right)^{r-i}  \left(\boldsymbol{\alpha}^T  G^{-1} \boldsymbol{c}_{0,0,0}^{(2i-2)} \right),    \label{Pen_expression_even}
\end{align}
where we have considered $(\boldsymbol{c}_0^{(0,0,0)})^T=(0,0,0,0)$.
Since we have obtained the analytical expression of $c^{(2r)}_{0,0,0,0}$,  we will consider
how to obtain all expansion coefficients $c_{i,j,k,l}^{(m)}$ by using the recurrence relation \eref{Pen-Next-Term}, \eref{Pen-Even-0000Term} and \eref{Pen-Odd-FirstTerm} step by step.
First we note \eref{Box-Odd-FirstTerm}  can be rewritten as
\begin{align}
\left(\begin{array}{l}
c^{(2r+1)}_{1,0,0,0} \\
c^{(2r+1)}_{0,1,0,0}\\
c^{(2r+1)}_{0,0,1,0} \\
c^{(2r+1)}_{0,0,0,1} \\
\end{array} \right) = (2r+1)   G^{-1} \left( c_{0,0,0,0}^{(2r)}~  \boldsymbol{\alpha}  -  \boldsymbol{c}_{0,0,0}^{(2r)} \right),   \label{Pen_expression_odd}
\end{align}
Since the expression of $c_{0,0,0,0}^{(2r)}$ and $\boldsymbol{c}_{0,0,0}^{(2r)}$ are known, then the results of $c_{1,0,0,0}^{(2r+1)}$, $c_{0,1,0,0}^{(2r+1)}$,$c_{0,0,1,0}^{(2r+1)}$, $c_{0,0,0,1}^{(2r+1)}$ are easy to get. Second, we take several examples to illustrate the procedure of calculation.
\begin{itemize}
	\item $m<4$: The essence of one-loop reduction is to expand the auxiliary vector $R$ with external momenta to cancel the propagators. To reduce a tensor pentagon integral to the scalar tadpole integral with propagator $P_0$, we need to cancel other four propagators, ie., $P_1,P_2,P_3,P_4$, which means four $R$'s are required at least.  So all the expansion coefficients vanish for rank $m<4$.
	\item $m=4$: We need to calculate $c^{(4)}_{0,0,0,0}$, $c^{(4)}_{4,0,0,0}$,$c^{(4)}_{2,2,0,0}$ and $c^{(4)}_{1,3,0,0}$,  other expansion coefficients can be got using the permutation symmetry.
	Using \eref{Pen-Even-0000Term}, we have
	\bea
	c_{0,0,0,0}^{(4)}=0.
	\eea
	Then using \eref{Pen-Next-Term}, we have
	\bea
	c^{(4)}_{1,1,0,0}&=&{1\over \Delta(1,2,3,4)}\Big[\Delta^{(4)}_{1,1}O_1^{(4)}(0,1,0,0)+\Delta^{(4)}_{1,2}O_2^{(4)}(0,1,0,0)\nn
	&&+\Delta^{(4)}_{1,3}O_3^{(4)}(0,1,0,0)+\Delta^{(4)}_{1,4}O_4^{(4)}(0,1,0,0)\Big]=0,\nn
	c^{(4)}_{2,0,0,0}&=&{1\over 2\Delta(1,2,3,4)}\Big[\Delta^{(4)}_{1,1}O_1^{(4)}(1,0,0,0)+\Delta^{(4)}_{1,2}O_2^{(4)}(1,0,0,0)\nn
	&&+\Delta^{(4)}_{1,3}O_3^{(4)}(1,0,0,0)+\Delta^{(4)}_{1,4}O_4^{(4)}(1,0,0,0)\Big]=0,
	\eea
	\bea
	c^{(4)}_{4,0,0,0}&=&{1\over 4\Delta(1,2,3,4)}\Big[\Delta^{(4)}_{1,1}O_1^{(4)}(3,0,0,0)+\Delta^{(4)}_{1,2}O_2^{(4)}(3,0,0,0)\nn
	&&+\Delta^{(4)}_{1,3}O_3^{(4)}(3,0,0,0)+\Delta^{(4)}_{1,4}O_4^{(4)}(3,0,0,0)\Big]\nn
	&=& {\beta_{12}\Delta(1,2;2,3)\over \beta_{11}\Delta(1,2)\Delta(1,2,3)}+[5\ permutations\ of\ (2,3,4)],
	\eea
	\bea
	c^{(4)}_{2,2,0,0}&=&{1\over  2\Delta(1,2,3,4)}\Big[\Delta^{(4)}_{1,1}O_1^{(4)}(1,2,0,0)+\Delta^{(4)}_{1,2}O_2^{(4)}(1,2,0,0)\nn
	&&+\Delta^{(4)}_{1,3}O_3^{(4)}(1,2,0,0)+\Delta^{(4)}_{1,4}O_4^{(4)}(1,2,0,0)\Big]\nn
	&=&{-6\Delta(2,3,4;1,2,4)\left(\beta _{24} \Delta (1,2)+\beta _{12} \Delta (2,4;1,2)\right)\over\beta _{22} \Delta (1,2) \Delta (1,2,4) \Delta(1,2,3,4)}+(3\leftrightarrow 4),
	\eea
	\bea
	c^{(4)}_{2,1,1,0}&=&{1\over  2\Delta(1,2,3,4)}\Big[\Delta^{(4)}_{1,1}O_1^{(4)}(1,1,1,0)+\Delta^{(4)}_{1,2}O_2^{(4)}(1,1,1,0)\nn
	&&+\Delta^{(4)}_{1,3}O_3^{(4)}(1,1,1,0)+\Delta^{(4)}_{1,4}O_4^{(4)}(1,1,1,0)\Big]\nn
	&=&\frac{12\Delta^{(4)}_{1,4}}{\Delta (1,2,3) \Delta (1,2,3,4)},
	\eea
	\bea
	c^{(4)}_{1,1,1,1}&=&{1\over  \Delta(1,2,3,4)}\Big[\Delta^{(4)}_{1,1}O_1^{(4)}(0,1,1,1)+\Delta^{(4)}_{1,2}O_2^{(4)}(0,1,1,1)\nn
	&&+\Delta^{(4)}_{1,3}O_3^{(4)}(0,1,1,1)+\Delta^{(4)}_{1,4}O_4^{(4)}(0,1,1,1)\Big]\nn
	&=&\frac{24}{ \Delta (1,2,3,4)},
	\eea
	\bea
	c^{(4)}_{1,3,0,0}&=&{1\over \Delta(1,2,3,4)}\Big[\Delta^{(4)}_{1,1}O_1^{(4)}(0,3,0,0)+\Delta^{(4)}_{1,2}O_2^{(4)}(0,3,0,0)\nn
	&&+\Delta^{(4)}_{1,3}O_3^{(4)}(0,3,0,0)+\Delta^{(4)}_{1,4}O_4^{(4)}(0,3,0,0)\Big]\nn
	&=&{-4\over \beta_{22}\Delta(1,2,3,4)}\Bigg\{\Bigg [\frac{ \Delta(2,3,4;1,2,4)\big  (\beta _{12} \Delta (2,4) \Delta (4,1;1,2)+\beta _{24} \Delta (1,2) \Delta (4,1;2,4)\big )}{ \Delta (1,2) \Delta (2,4) \Delta (1,2,4)}\nn
	&&+\frac{\beta _{23} \Delta (2,4) \Delta (3,4;2,3)}{ \Delta (2,3) \Delta (2,4) }\Bigg]+(3\leftrightarrow 4)\Bigg\},
	\eea
	where we have used
	\bea
	O_1^{(4)}(1,0,0,0)&=&4\a_1c^{(3)}_{1,0,0,0}-4c^{(4)}_{0,0,0,0}=0,\nn
	O_2^{(4)}(1,0,0,0)&=&4\left(\a_2c^{(3)}_{1,0,0,0}-c^{(3)}_{1,0,0}[\hat{2}]\right)=0,\nn
	O_3^{(4)}(1,0,0,0)&=&4\left(\a_3c^{(3)}_{1,0,0,0}-c^{(3)}_{1,0,0}[\hat{3}]\right)=0,\nn
	O_4^{(4)}(1,0,0,0)&=&4\left(\a_4c^{(3)}_{1,0,0,0}-c^{(3)}_{1,0,0}[\hat{4}]\right)=0,\no
	\eea
	\bea
	O_1^{(4)}(0,1,0,0)&=&4\left(\a_1c^{(3)}_{0,1,0,0}-c^{(3)}_{1,0,0}[\hat{1}]\right)=0,\nn
	O_2^{(4)}(0,1,0,0)&=&4\left(\a_2c^{(3)}_{0,1,0,0}\right)-4c^{(4)}_{0,0,0,0}=0,\nn
	O_3^{(4)}(0,1,0,0)&=&4\left(\a_3c^{(3)}_{0,1,0,0}-c^{(3)}_{0,0,0}[\hat{3}]\right)=0,\nn
	O_4^{(4)}(0,1,0,0)&=&4\left(\a_4c^{(3)}_{0,1,0,0}-c^{(3)}_{0,0,0}[\hat{4}]\right)=0,\no
	\eea
	\bea
	O_1^{(4)}(3,0,0,0)&=&4\a_1c^{(3)}_{3,0,0,0}-2c^{(4)}_{2,0,0,0}=0,\no
	\eea
	\bea
	O_2^{(4)}(3,0,0,0)&=&4\left(\a_2c^{(3)}_{3,0,0,0}-c^{(3)}_{3,0,0}[\hat{2}]\right)\nn
	&=&-\frac{4 \left(\beta _{13} \Delta (1,4) \Delta (3,4;1,3)+\beta _{1,4} \Delta (1,3) \Delta (3,4;4,1)\right)}{\beta _{11} \Delta (1,3) \Delta (1,4) \Delta (1,3,4)},\no\eea
	\bea
	O_3^{(4)}(3,0,0,0)&=&4\left(\a_3c^{(3)}_{3,0,0,0}-c^{(3)}_{3,0,0}[\hat{3}]\right)\nn
	&=&-\frac{4 \left(\beta _{12} \Delta (1,4) \Delta (2,4;1,2)+\beta _{1,4} \Delta (1,2) \Delta (2,4;4,1)\right)}{\beta _{11} \Delta (1,2) \Delta (1,4) \Delta (1,2,4)},\no\eea
	\bea
	O_4^{(4)}(3,0,0,0)&=&4\left(\a_4c^{(3)}_{3,0,0,0}-c^{(3)}_{3,0,0}[\hat{4}]\right)\nn
	&=&-\frac{4 \left(\beta _{12} \Delta (1,3) \Delta (2,3;1,2)+\beta _{13} \Delta (1,2) \Delta (2,3;3,1)\right)}{\beta _{11} \Delta (1,2) \Delta (1,3) \Delta (1,2,3)},\no
	\eea
	\bea
	O_1^{(4)}(0,3,0,0)&=&4\left(\a_1c^{(3)}_{0,3,0,0}-c^{(3)}_{3,0,0}[\hat{1}]\right)\nn
	&=&-\frac{4 \left(\beta _{23} \Delta (2,4) \Delta (3,4;2,3)+\beta _{24} \Delta (2,3) \Delta (3,4;4,2)\right)}{\beta _{22} \Delta (2,3) \Delta (2,4) \Delta (2,3,4)},\nn
	O_2^{(4)}(0,3,0,0)&=&4\a_2c^{(3)}_{0,3,0,0}-2c^{(4)}_{0,2,0,0}=0,\nn
	O_3^{(4)}(0,3,0,0)&=&4\left(\a_3c^{(3)}_{0,3,0,0}-c^{(3)}_{0,0,0}[\hat{3}]\right)\nn
	&=&-\frac{4 \left(\beta _{12} \Delta (2,4) \Delta (4,1;1,2)+\beta _{24} \Delta (1,2) \Delta (4,1;2,4)\right)}{\beta _{22} \Delta (1,2) \Delta (2,4) \Delta (1,2,4)},\nn
	O_4^{(4)}(0,3,0,0)&=&4\left(\a_4c^{(3)}_{0,3,0,0}-c^{(3)}_{0,0,0}[\hat{4}]\right)\nn
	&=&-\frac{4 \left(\beta _{12} \Delta (2,3) \Delta (3,1;1,2)+\beta _{23} \Delta (1,2) \Delta (3,1;2,3)\right)}{\beta _{22} \Delta (1,2) \Delta (2,3) \Delta (1,2,3)},\no
	\eea
	\bea
	O_1^{(4)}(1,2,0,0)&=&4\a_1c^{(3)}_{1,2,0,0}-2c^{(4)}_{0,2,0,0}=0,\nn
	O_2^{(4)}(1,2,0,0)&=&4\a_2c^{(3)}_{1,2,0,0}-2c^{(4)}_{1,1,0,0}=0,\nn
	O_3^{(4)}(1,2,0,0)&=&4\left(\a_3c^{(3)}_{1,2,0,0}-c^{(3)}_{1,0,0}[\hat{3}]\right)=-\frac{12 \left(\beta _{24} \Delta (1,2)+\beta _{12} \Delta (2,4;1,2)\right)}{\beta _{22} \Delta (1,2) \Delta (1,2,4)},\nn
	O_4^{(4)}(1,2,0,0)&=&4\left(\a_4c^{(3)}_{1,2,0,0}-c^{(3)}_{1,0,0}[\hat{4}]\right)=-\frac{12 \left(\beta _{23} \Delta (1,2)+\beta _{12} \Delta (2,3;1,2)\right)}{\beta _{22} \Delta (1,2) \Delta (1,2,3)},\no
	\eea
	\bea
	O_1^{(4)}(1,1,1,0)&=&4\a_1c^{(3)}_{1,1,1,0}-4c^{(4)}_{0,1,1,0}=0,\nn
	O_2^{(4)}(1,1,1,0)&=&4\a_2c^{(3)}_{1,1,1,0}-4c^{(4)}_{1,0,1,0}=0,\nn
	O_3^{(4)}(1,1,1,0)&=&4\a_3c^{(3)}_{1,1,1,0}-4c^{(4)}_{1,1,0,0}=0,\nn
	O_4^{(4)}(1,1,1,0)&=&4\a_4c^{(3)}_{1,1,1,0}-4c^{(3)}_{1,1,1}[\hat{4}]=\frac{24}{\Delta (1,2,3)},\no
	\eea
	\bea
	O_1^{(4)}(0,1,1,1)&=&4\a_1c^{(3)}_{0,1,1,1}-c^{(3)}_{1,1,1}[\hat{1}]=\frac{24}{\Delta (2,3,4)},\nn
	O_2^{(4)}(0,1,1,1)&=&4\a_2c^{(3)}_{0,1,1,1}-4c^{(4)}_{0,0,1,1}=0,\nn
	O_3^{(4)}(0,1,1,1)&=&4\a_3c^{(3)}_{0,1,1,1}-4c^{(4)}_{0,1,0,1}=0,\nn
	O_4^{(4)}(0,1,1,1)&=&4\a_4c^{(3)}_{0,1,1,1}-4c^{(4)}_{0,1,1,0}=0.\no
	\eea
\end{itemize}
The above examples are enough to illustrate the procedure of calculating expansion coefficients $c_{i,j,k,l}^{(m)}$.
Rough speaking, we calculate the expansion coefficients $c_{i,j,k,l}^{(m)}$ from lower rank to higher rank, and at a fixed rank $m$, we prefer to calculate the expansion coefficients $c_{i,j,k,l}^{(m)}$ with smaller index $i+j+k+l$ first. For general rank $m_0$, we need to calculate all expansion coefficients with rank $m\le m_0$ to calculate the tadpole coefficient with rank $m_0$. Now our algorithm of the calculating tadpole coefficient with rank $m_0$ is summarized as following
\footnote{ It is similar to the tensor triangle, but due to there are four indices in $c_{i,j,k,l}^{(m)}$, we don't draw a picture to illustrate the algorithm.}
\begin{itemize}
	
	\item Step 1: For rank $m<4$, we have shown that all expansion coefficients vanish.
	
	\item Step 2: Consider the rank $m=4$, calculate the expansion coefficients $c^{(4)}_{0,0,0,0}, c^{(4)}_{2,0,0,0}$, $c^{(4)}_{1,1,0,0}$,
	$\cdots,c^{(4)}_{4,0,0,0}$  by  and \eref{Pen-Even-0000Term} and \eref{Pen-Next-Term} as illustrated before.
	
	\item Step 3: Consider the rank $m=5$,  Combining the permutation symmetry, calculate the expansion coefficients $c^{(5)}_{1,0,0,0}, c^{(5)}_{1,1,1,0},\cdots,c^{(5)}_{5,0,0,0}$ successively by \eref{Pen-Even-0000Term} and \eref{Pen-Next-Term} as illustrated before.
	
	$\cdots$
	
	\item Step $m \le m_0$: Combine the permutation symmetry,
	if $m=2r$, calculate $c_{0,0,0,0}^{(2r)}$,$ c_{1,1,0,0}^{(2r)}$,\\$c_{2,0,0,0}^{(2r)}$,$\cdots$,$c_{2r,0,0,0}^{(2r)}$ successively \eref{Pen-Even-0000Term} and \eref{Pen-Next-Term}.
	if $m=2r+1$, calculate $c_{1,0,0,0}^{(2r+1)}, c_{1,1,1,0}^{(2r+1)}$,\\$c_{1,2,0,0}^{(2r+1)},c_{3,0,0,0}^{(2r+1)}, \cdots,c_{2r+1,0,0,0}^{(2r+1)}$ successively by using \eref{Pen-Next-Term}.
	\item Final step: Combine all expansion coefficients to get the tadpole coefficient by \eref{Pen-Exp}.
\end{itemize}

With the help of Mathematica, it is easy to implement recurrence relations \eref{Pen_expression_even} and \eref{Pen-Next-Term-Matrix} to automatically
generate analytic expression of reduction coefficients of any rank.

% % % % % % % % % % % % % %
% % % % % % % % % % % % % %
\section{Conclusion}
%%%%%%%%%%%%%%%%%%%%%%
In this paper, we have considered calculating the reduction tadpole coefficient of general one-loop integrals. This piece is missed part in the standard unitarity cut method. By introducing 
the auxiliary vector $R$ and the trick of differentiation over the auxiliary vector $R$,
we get the differential equations for tadpole coefficients. By  expanding the tadpole coefficients
according to its tensor structure, we get some recurrence relations for expansion coefficients. 
It is easy to organize recurrence relations to the form that coefficients of higher rank and higher indices are expressed by coefficients of lower rank and lower indices, which can easily be implemented into Mathematica and gives the expression of tadpole coefficients automatically.

%Except the tensor tadpole, these recurrence relations are hard to be solved analytically. In practice, we combine these recurrence relations and the boundary conditions to calculate the expansion coefficients from lower-point integral to higher-point integral and from lower rank to high rank. For general case, we have presented a simple and effective algorithm, which can work in Mathematica and gives the expression of tadpoles automatically. In practice, our algorithm is quite fast.

To demonstrate our algorithm, after discussing the recurrence relations for tensor bubbles, triangles, boxes and pentagons, we have shown the calculation of some examples. Moreover, by  changing the boundary conditions, our algorithm can be applied to calculating the reduction coefficients of other master integrals. We will show how to do this in the further research.

As for the reduction of tensor higher-loop integral, by constructing differential operators and expanding the coefficients in a general form, our method can also give the recurrence relations.  But differing from one-loop case, these relations are in general not enough to uniquely determine coefficients. A further complexity is that for the higher-loop integrals, the master basis are more complicated. In despite of these difficulties, it is still an interesting question to apply our method to  higher-loop cases.

% % % % % % % % % %
\section*{Acknowledgments}
% % % % % % % % % %
It is a pleasure to thank Yang Zhang, Chang Hu and Yaobo Zhang for inspiring discussions. This work is supported by
Qiu-Shi Funding and Chinese NSF funding under Grant No.11935013, No.11947301, No.12047502 (Peng
Huanwu Center).
\appendix

%%%%%%%%%%%%%%%%%%
\section{The reduction of tadpole by PV-method}
\label{sec:PVreduction}
%%%%%%%%%%%%%%%%%%

In this appendix, we use the traditional PV-reduction method to study following tadpole integrals:
\bea {\cal A}^{\mu_1 ...\mu_s}(M_0) & = & \frac{1}{i \pi^{D/2}}\int \; d^D \ell  \; \frac{\ell^{\mu_1}...
\ell^{\mu_s}} {P_0},~~~~P_0=\ell^2-M_0^2,~~~~\label{Tadpole-tensor-1-1}\eea
The key of PV-reduction is that  by Lorentz symmetry  the tensor structure at the both sides of \eref{Tadpole-tensor-1-1} must be the same. For the tadpole \eref{Tadpole-tensor-1-1}, there is no external momentum to provide the tensor index, thus the only available one is  the metric $g^{\mu\nu}$, which must be considered  in the general $D=4-2\eps$-dimension.
Furthermore, because all $\mu_i$'s are symmetric, the tensor structure must be symmetric under index permutation.  Thus we have when $s=2r+1$, \eref{Tadpole-tensor-1-1} is zero and when $s=2r$ we have
\bea {\cal A}^{\mu_1 ...\mu_{2r}}(M_0)=A(2r) [g^{\mu_1 \mu_2}...g^{\mu_{2r-1}\mu_{2r}}+{\rm symmetrization}] {\cal A}^{s=0}(M_0).~~~~\label{Tadpole-tensor-1-2}\eea
To determine the constant $A(2r)$, we contract both sides of \eref{Tadpole-tensor-1-1} with, for example, $g_{\mu_1\mu_2}$. Using \eref{Tadpole-tensor-1-2} the RHS gives
\bea A(2r)(D+2(r-1)) [  g^{\mu_3 \mu_4}...g^{\mu_{2r-1}\mu_{2r}}+ {\rm symmetrization} ],~r\geq 2.~~~~\label{Tadpole-tensor-1-3}\eea
 When doing the contraction, there are two types of
tensor structures in \eref{Tadpole-tensor-1-2}, one is with $g^{\mu_1\mu_2}$ and another one, with $g^{\mu_1\mu_i} g^{\mu_2\mu_j}$. For the former, $g_{\mu_1\mu_2}g^{\mu_1\mu_2}=D$, while for the later $g_{\mu_1\mu_2}g^{\mu_1\mu_i} g^{\mu_2\mu_j}=g^{\mu_i \mu_j}$. Furthermore, for the later one, since for each $i,j$ pair, which are different tensor structures before contraction,
we get the same tensor structure after the contraction, thus when considering the remaining particular tensor structure, for example,
$g^{\mu_3\mu_4}... g^{\mu_{2r-1}\mu_{2r}}$ with $(r-1)$'s $g^{\mu\nu}$, we get the overall factor  $2(r-1)$.

For the LHS, we have
\bea & & \frac{1}{i \pi^{D/2}}\int \; d^D \ell \; \frac{\ell^2 \ell^{\mu_3}...
\ell^{\mu_s}} {\ell^2-M_0^2+i\eps}=M_0^2\frac{1}{i \pi^{D/2}}\int \; d^D \ell  \; \frac{ \ell^{\mu_3}...
\ell^{\mu_s}} {\ell^2-M_0^2+i\eps}+\frac{1}{i \pi^{D/2}}\int \; d^D \ell  \;  \ell^{\mu_3}...
\ell^{\mu_s} \nn
	& = & M	_0^2 A(2r-2) [  g^{\mu_3 \mu_4}...g^{\mu_{2r-1}\mu_{2r}}+ {\rm symmetrization} ], ~~~~\label{Tadpole-tensor-1-4}\eea
where the second term in the first line belongs to the type of {\bf scaleless integral}, which is zero by definition under dimensional regularization scheme.
Comparing these two calculations, we get the recurrence relation
\bea A(2r)= {M_0^2\over D+2(r-1)} A(2r-2),~~~~\label{Tadpole-tensor-1-6}\eea
Using the boundary condition $A(0)=1$, we get\footnote{One can check this result with
Peskin's book,(A.44)-- (A.48) where one put $n=1$ \cite{Peskin:1995ev}. Another checking can be found in \cite{Denner:2005nn}, (3.1)---(3.4): one can check that when $D=4$,
$\prod_{t=1}^{r}(D+2(t-1))=2^r (r+1)!$. }
\bea A(2r)= {M_0^{2r}\over \prod_{t=1}^{r}(D+2(t-1))},~~~r\geq 1.~~~~\label{Tadpole-tensor-1-7}\eea
If we contract with $R$ in \eref{Tadpole-tensor-1-1}, using \eref{Tadpole-tensor-1-2} we get zero if $s$ is odd and
\bea I_{s,1}= A(s=2r) \kappa (R^2)^{r}\int {d^D\ell\over (2\pi)^D} { 1\over (\ell^2-M_0^2)},~~~~\label{Tadpole-tensor-1-8} \eea
if $s$ is even where the $\kappa$ is the total number of tensor structures in \eref{Tadpole-tensor-1-2} given by
\bea \kappa={(2r)!\over 2^r r!}.~~~~\label{Tadpole-tensor-1-9}\eea
One can check that coefficients in \eref{Tadpole-tensor-1-8} is the same as the one given in \eref{tad-coeff-3}.
\bibliographystyle{JHEP}
\bibliography{reference}

\end{document}